\documentclass[pra,twocolumn,superscriptaddress,preprintnumbers,showpacs]{revtex4-1}
\usepackage{amsmath,amssymb,bm,graphicx,hyperref}

\renewcommand\d{\partial}
\newcommand\grad{\bm{\nabla}}
\newcommand\+{\dagger}
\newcommand\<{\langle}
\renewcommand\>{\rangle}
\renewcommand\r{{\bm{r}}}

\renewcommand\k{{\bm{k}}}
\newcommand\0{{\bm{0}}}
\newcommand\n{{\bm{n}}}

\newcommand\eff{{\mathrm{eff}}}
\newcommand\G{\mathcal{G}}

\begin{document}
\preprint{MIT-CTP 4184}

\title{Confinement-induced $p$-wave resonances from $s$-wave interactions}
\author{Yusuke~Nishida}
\affiliation{Center for Theoretical Physics,
Massachusetts Institute of Technology, Cambridge, Massachusetts 02139, USA}
\author{Shina~Tan}
\affiliation{Department of Physics, Yale University,
New Haven, Connecticut 06520, USA}
\affiliation{School of Physics, Georgia Institute of Technology,
Atlanta, Georgia 30332, USA}

\begin{abstract}
 We show that a purely $s$-wave interaction in three dimensions (3D) can
 induce higher partial-wave resonances in mixed dimensions.  We develop
 two-body scattering theories in all three cases of 0D-3D, 1D-3D, and
 2D-3D mixtures and determine the positions of higher partial-wave
 resonances in terms of the 3D $s$-wave scattering length assuming a
 harmonic confinement potential.  We also compute the low-energy
 scattering parameters in the $p$-wave channel (scattering volume and
 effective momentum) that are necessary for the low-energy effective
 theory of the $p$-wave resonance.  We point out that some of the
 resonances observed in the Florence group experiment
 [Phys.\ Rev.\ Lett.\ {\bf 104}, 153202 (2010)] can be interpreted as
 the $p$-wave resonances in the 2D-3D mixed dimensions.  Our study paves
 the way for a variety of physics, such as Anderson localization of
 matter waves under $p$-wave resonant scatterers.
\end{abstract}

\date{October 2010}


\maketitle

\section{Introduction \label{sec:introduction}}
Scattering resonances play a central role in cold-atom
physics~\cite{Chin:2010}.  In particular, $s$-wave Feshbach resonances
induced by a magnetic field have been utilized to control the atom-atom
interaction and have led to experimental realization of a rich variety
of physics, such as the BCS-BEC crossover in Fermi
gases~\cite{Ketterle:2008,Bloch:2008,Giorgini:2008}.  Although a
$p$-wave analog of the BCS-BEC crossover has been predicted
theoretically~\cite{Ho:2005,Botelho:2005,Ohashi:2005,Gurarie:2005,Cheng:2005}
and $p$-wave and higher partial-wave resonances have been observed
experimentally~\cite{Regal:2003,Zhang:2004,Schunck:2005,Chevy:2005,Gunter:2005,Gaebler:2007,Jin:2008,Fuchs:2008,Inada:2008,Maier:2010,Chin:2010},
$p$-wave superfluids have not been realized in cold-atom experiments so
far.  This is because Fermi gases in the vicinity of the $p$-wave
Feshbach resonances are unstable owing to inelastic collisions decaying
into deeply bound dimers and do not reach their equilibrium within their
short lifetime~\cite{Levinsen:2007,Jona-Lasinio:2008}.  In contrast,
Fermi gases in the vicinity of the $s$-wave Feshbach resonances are
long lived because such inelastic collisions are strongly suppressed by
the Pauli exclusion principle~\cite{Petrov:2003}.

\begin{figure}[b]
 \includegraphics[width=\columnwidth,clip]{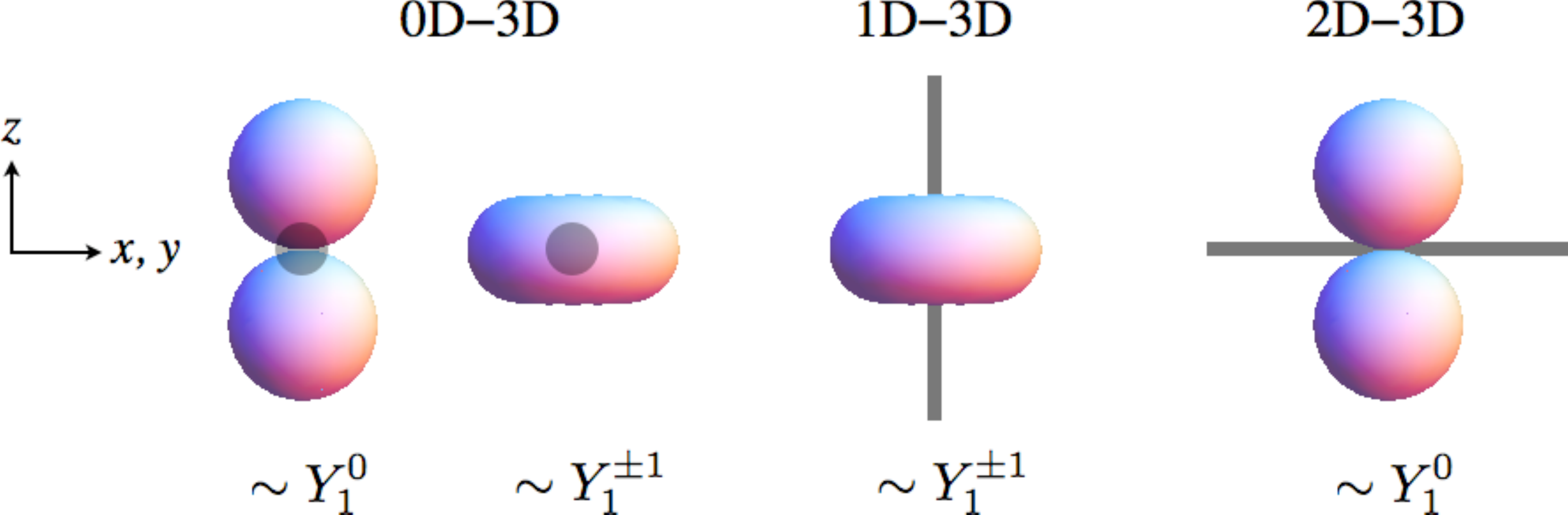}
 \caption{(Color online) Illustrations of shallow $p$-wave molecules in
 mixed dimensions.  Shown are the asymptotic angular wave functions of
 $B$ atoms in 3D relative to $A$ atoms confined in 0D (origin), 1D
 ($z$ axis), and 2D ($xy$ plane), respectively.  See the text for
 details.  \label{fig:molecule}}
\end{figure}

In this paper, we propose a way to induce $p$-wave and higher
partial-wave resonances from the $s$-wave Feshbach resonance only.  This
can be achieved by using a mixture of two atomic species $A$ and $B$ in
mixed dimensions, where $A$ atoms are confined in lower dimensions
[quasi-zero dimension (0D), one dimension (1D), or two dimensions (2D)]
and interact with $B$ atoms living in three dimensions
(3D)~\cite{Massignan:2006,Nishida:2008kr} (see Fig.~\ref{fig:molecule}).
Indeed, the authors of Ref.~\cite{Massignan:2006} mention the possibility
of $p$-wave resonances in the 0D-3D mixture.  Higher partial-wave
resonances induced from a purely $s$-wave interaction seem
counterintuitive but can be understood by generalizing the argument for
$s$-wave resonances given in Ref.~\cite{Massignan:2006}.

Suppose an $s$-wave interspecies interaction between $A$ and $B$ atoms
supports a bound state with its binding energy $E_b<0$.  We confine only
the $A$ atom by a 3D harmonic potential with the oscillator frequency
$\omega$.  Because the $AB$ molecule also feels the confinement
potential, its center-of-mass motion is quantized.  In the limit
$|E_b|\gg\hbar\omega$ or $m_A\gg m_B$, the energy of the $AB$ molecule
$E_{AB}$ is given by a sum of its center-of-mass energy and binding
energy:
\begin{equation}\label{eq:approx_0D}
 E_{AB} = \left(\frac32+\ell+2n\right)\sqrt\frac{m_A}{m_A+m_B}\,\hbar\omega-|E_b|,
\end{equation}
where $m_A\,(m_B)$ is the mass of the $A\,(B)$ atom, $\ell=0,1,2,\dots$
is an orbital angular momentum quantum number, and $n\geq0$ is an
integer.  We can view these quantized energy levels as a tower of
$\ell$th partial-wave ``Feshbach molecules.''  When one of them
coincides with the $A$-$B$ scattering threshold at $\frac32\hbar\omega$,
the coupling between the center-of-mass and relative motions leads to
the resonance occurring in the $\ell$th partial-wave channel.
Furthermore, there exists a series of resonances in each partial-wave
channel corresponding to different values of $n$.

Similarly, in the 1D-3D mixture, the $A$ atom is confined by a 2D
harmonic potential and the energy of the $AB$ molecule is quantized as
\begin{equation}\label{eq:approx_1D}
 E_{AB} = \left(1+|m|+2n\right)\sqrt\frac{m_A}{m_A+m_B}\,\hbar\omega-|E_b|,
\end{equation}
where $m=0,\pm1,\pm2,\dots$ is a magnetic quantum number.  When one of
them coincides with the $A$-$B$ scattering threshold at $\hbar\omega$,
the resonance occurs in the $|m|$th partial-wave channel.  There exists
a series of resonances in each partial-wave channel corresponding to
different values of $n$.

In the 2D-3D mixture, the $A$ atom is confined by a 1D harmonic
potential and the energy of the $AB$ molecule is quantized as
\begin{equation}\label{eq:approx_2D}
 E_{AB} = \left(\frac12+n\right)\sqrt\frac{m_A}{m_A+m_B}\,\hbar\omega-|E_b|.
\end{equation}
When one of them coincides with the $A$-$B$ scattering threshold at
$\frac12\hbar\omega$, the resonance occurs.  Because the wave function
of the $AB$ molecule has the even (odd) parity for an even (odd) $n$,
the associated resonance is in the even (odd)-parity channel, which is
also phrased as an $s$-wave ($p$-wave) resonance in the 2D-3D mixture.
There exists a series of resonances in each parity channel corresponding
to different values of $n$.  In order to achieve these higher
partial-wave resonances in mixed dimensions, one needs to apply a strong
optical lattice only on $A$ atoms without affecting $B$
atoms~\cite{LeBlanc:2007}.  Such a species-selective optical lattice has
been successfully implemented in a Bose-Bose mixture of $^{41}$K and
$^{87}$Rb~\cite{Catani:2009} and the 2D-3D mixed dimensions have been
realized in the Florence group experiment~\cite{Lamporesi:2010}.

The above simple arguments illustrate how the higher partial-wave
resonances are induced in mixed dimensions from the purely $s$-wave
interaction in a free space.  In this paper, we develop two-body
scattering theories in mixed dimensions, assuming the harmonic
confinement potential and the short-range interspecies interaction
characterized by an $s$-wave scattering length ($\equiv a_\mathrm{3D}$)
and effective range ($\equiv r_\mathrm{3D}$), which are extensions of
analyses in Refs.~\cite{Massignan:2006,Nishida:2008kr} aimed for
$s$-wave scatterings only.  The two-body scattering of a 0D, 1D, or 2D
atom with a 3D atom is three-dimensional in the sense that it is
described by three relative coordinates, and thus such scattering
theories can be developed in parallel with the ordinary scattering
theory in 3D.  In particular, we determine the positions of all
partial-wave resonances and will see that they are well understood from
the above simple arguments in a wide range of the mass ratio
$m_A/m_B\gtrsim1$.  We also compute the low-energy scattering parameters
in the $p$-wave channel (scattering volume and effective momentum) that
are necessary for the low-energy effective theory of the $p$-wave
resonance in mixed dimensions.

Our analyses and results for the 0D-3D, 1D-3D, and 2D-3D mixtures are
presented in Secs.~\ref{sec:0D-3D}, \ref{sec:1D-3D}, and
\ref{sec:2D-3D}, respectively, while some details are shown in
Appendixes~\ref{app:0D-3D}, \ref{app:1D-3D}, and \ref{app:2D-3D},
respectively.  Also in Sec.~\ref{sec:2D-3D}, we point out that some of
the resonances observed in the Florence group
experiment~\cite{Lamporesi:2010} can be interpreted as the $p$-wave
resonances in the 2D-3D mixed dimensions.  Finally,
Sec.~\ref{sec:summary} is devoted to a summary of this paper and the
stability of confinement-induced molecules and their analogy with
Kaluza-Klein modes in extra-dimension models are discussed.

For later use, we define the reduced mass
$m_{AB}\equiv m_Am_B/(m_A+m_B)$, the total mass $M\equiv m_A+m_B$, and
the harmonic oscillator length
$l_\mathrm{ho}\equiv\sqrt{\hbar/m_A\omega}$.  Below we set $\hbar=1$ and
the range of integrations is assumed to be from $-\infty$ to $\infty$
unless otherwise specified.

\section{0D-3D mixed dimensions \label{sec:0D-3D}}

\subsection{Scattering theory~\cite{Massignan:2006}}
The scattering of a quasi-0D $A$ atom with a $B$ atom in 3D is described
by a Schr\"odinger equation,
\begin{equation}\label{eq:schrodinger_0D}
 \begin{split}
  & \left(-\frac{\grad_{\!\r_A}^2}{2m_A}+\frac12m_A\omega^2\r_A^2
  -\frac{\grad_{\!\r_B}^2}{2m_B}\right)\psi(\r_A,\r_B) \\
  &= E\,\psi(\r_A,\r_B)
 \end{split}
\end{equation}
for $|\r_A-\r_B|>0$.  The short-range interspecies interaction is
implemented by the generalized Bethe-Peierls boundary
condition~\cite{Petrov:2004,Levinsen:2009mn}:
\begin{equation}\label{eq:short-range_0D}
 \psi(\r_A,\r_B)\big|_{\r_A,\r_B\to\r} \to 
  \left[\frac1{\tilde a(\hat E_c)}-\frac1{|\r_A-\r_B|}\right]\chi(\r).
\end{equation}
Here $\tilde a(\hat E_c)$ is the energy-dependent ``scattering length''
defined by
\begin{equation}\label{eq:generalized_a}
 \frac1{\tilde a(\hat E_c)} \equiv 
  \frac1{a_\mathrm{3D}} - m_{AB}r_\mathrm{3D}\hat E_c
\end{equation}
and the collision energy operator $\hat E_c$ in the present case is
given by
\begin{equation}
 \hat E_c = E
  - \left(-\frac{\grad_{\!\r}^2}{2M}+\frac12m_A\omega^2\r^2\right).
\end{equation}

The solution to the Schr\"odinger equation (\ref{eq:schrodinger_0D}) can
be written as
\begin{equation}\label{eq:solution_0D}
 \begin{split}
  & \psi(\r_A,\r_B) = \psi_0(\r_A,\r_B) \\
  &\quad + \frac{2\pi}{m_{AB}}\int\!d\r'G_E(\r_A,\r_B;\r',\r')\chi(\r'),
 \end{split}
\end{equation}
where $\psi_0$ is a solution in the noninteracting limit and $G_E$ is the
retarded Green's function for the noninteracting Hamiltonian:
\begin{equation}\label{eq:G_0D}
 \begin{split}
  & G_E(\r_A,\r_B;\r'_A,\r'_B) \\
  &\equiv \<\r_A,\r_B|\frac1{E-H_0+i0^+}|\r'_A,\r'_B\> \\
  &= -\frac{m_B}{2\pi}\sum_\n\phi_\n(\r_A)\phi_\n^*(\r'_A) \\
  &\quad \times \frac{e^{-\sqrt{2m_B}\sqrt{(n_x+n_y+n_z+\frac32)\omega-E-i0^+}
  |\r_B-\r_B'|}}{|\r_B-\r'_B|}.
 \end{split}
\end{equation}
Here $\phi_\n$ with quantum numbers $\n=(n_x,n_y,n_z)$ is the normalized
wave function of an $A$ atom in the 3D harmonic potential.

We now consider the low-energy scattering in which
\begin{equation}
 E-\frac32\omega \equiv \frac{k^2}{2m_B} \ll \omega
\end{equation}
is satisfied, and then $\psi_0$ becomes
\begin{equation}
 \psi_0(\r_A,\r_B) = Ce^{i\k\cdot\r_B}\phi_\0(r_A),
\end{equation}
which represents the $A$ atom in the ground state of the 3D harmonic
potential and the plane wave of $B$ atom with the wave vector $\k$.  The
asymptotic form of the wave function at a large distance
$|\r_B|\gg l_\mathrm{ho}$ is given by
\begin{equation}\label{eq:asymptotic_0D}
 \psi(\r_A,\r_B) \to C\left[e^{i\k\cdot\r_B}
  +\frac{e^{ikr_B}}{r_B}f(\k,\k')\right]\phi_\0(r_A),
\end{equation}
where $f(\k,\k')$ with $\k'\equiv k\hat\r_B$ defines the two-body
scattering amplitude in the 0D-3D mixed dimensions:
\begin{equation}\label{eq:f_0D}
 f(\k,\k') \equiv -\frac1{C}\frac{m_B}{m_{AB}}
  \int\!d\r'e^{-i\k'\cdot\r'}\phi_\0^*(r')\chi(\r').
\end{equation}
We note that $\chi$ has an implicit $\k$ dependence.

The unknown function $\chi$ can be determined by substituting the
solution (\ref{eq:solution_0D}) into the Bethe-Peierls boundary
condition (\ref{eq:short-range_0D}).  Defining the regular part of the
Green's function $\G$ by
\begin{equation}\label{eq:regular_0D}
 \begin{split}
  & G_E(\r_A,\r_B;\r',\r')\big|_{\r_A,\r_B\to\r} \\
  &\equiv -\frac{m_{AB}}{2\pi|\r_A-\r_B|}\delta(\r-\r') + \G(\r;\r'),
 \end{split}
\end{equation}
we obtain
\begin{equation}\label{eq:chi_0D}
 \frac1{\tilde a(\hat E_c)}\chi(\r) = C e^{i\k\cdot\r}\phi_\0(r)
  + \frac{2\pi}{m_{AB}}\int\!d\r'\G(\r;\r')\chi(\r').
\end{equation}
This integral equation determines $\chi/C$, which in turn provides $f$
from Eq.~(\ref{eq:f_0D}).

Because of the 3D rotational symmetries of the system, $f$, $\chi$, and
$\G$ can be decomposed into their partial-wave components:
\begin{align}\label{eq:decomposition_0D}
 f(\k,\k') &= \sum_{\ell=0}^\infty\left(2\ell+1\right)
 f_\ell(k)\,P_\ell(\hat\k\cdot\hat\k'), \\
 \chi(\r) &= \sum_{\ell=0}^\infty\left(2\ell+1\right)
 \chi_\ell(r)\,P_\ell(\hat\r\cdot\hat\k), \\
 \G(\r;\r') &= \sum_{\ell=0}^\infty\left(2\ell+1\right)
 \G_\ell(r;r')\,P_\ell(\hat\r\cdot\hat\r').
\end{align}
Equations~(\ref{eq:f_0D}) and (\ref{eq:chi_0D}) lead to the $\ell$th
partial-wave scattering amplitude given by
\begin{equation}\label{eq:partial-wave_0D}
 f_\ell(k) = -\frac1{C}\frac{m_B}{m_{AB}}
  \int\!d\r'j_\ell(kr')\phi_\0^*(r')\chi_\ell(r')
\end{equation}
with
\begin{equation}
 \begin{split}
  \frac1{\tilde a(\hat E_c)}\chi_\ell(r) &= Cj_\ell(kr)\phi_\0(r) \\
  &\quad + \frac{2\pi}{m_{AB}}\int\!d\r'\G_\ell(r;r')\chi_\ell(r').
 \end{split}
\end{equation}
From the general argument based on the 3D rotational symmetries and
unitarity~\cite{Sakurai}, or from the explicit calculation that uses the
Green's function in Eq.~(\ref{eq:G_0D}), we can show that $f_\ell$ has
the usual low-energy expansion:
\begin{equation}\label{eq:expansion_0D}
 \lim_{k\to0}f_\ell(k) = -\frac{k^{2\ell}}
  {\frac1{a_\eff^{(\ell)}}-\frac12r_\eff^{(\ell)}k^2+O(k^4)+ik^{2\ell+1}},
\end{equation}
where $a_\eff^{(\ell)}$ and $r_\eff^{(\ell)}$ are effective scattering
``length'' and ``range'' parameters in the $\ell$th partial-wave
channel.  Note that $a_\eff^{(\ell)}$ has the dimension of
$(\mathrm{length})^{2\ell+1}$ and $r_\eff^{(\ell)}$ has
$(\mathrm{length})^{1-2\ell}$.

Substituting the expansion of $f_\ell$ into
Eq.~(\ref{eq:partial-wave_0D}), we can determine the low-energy
scattering parameters.  In particular, the effective scattering length
$a_\eff^{(\ell)}$ is given by
\begin{equation}\label{eq:a_eff_0D-1}
 a_\eff^{(\ell)} = \frac1{C}\frac{m_B}{m_{AB}}
  \frac{\sqrt\pi}{2^{\ell+1}\left(\ell+\frac12\right)!}
  \int\!d\r'r'^\ell\phi_\0^*(r')\chi_\ell(r')
\end{equation}
with
\begin{equation}\label{eq:a_eff_0D-2}
 \begin{split}
  \frac1{\tilde a(\hat E_c)}\chi_\ell(r) 
  &= C\frac{\sqrt\pi}{2^{\ell+1}\left(\ell+\frac12\right)!}r^\ell\phi_\0(r) \\
  &\quad + \frac{2\pi}{m_{AB}}\int\!d\r'\G_\ell(r;r')\big|_{k\to0}\chi_\ell(r').
 \end{split}
\end{equation}
The $\ell$th partial-wave resonance in the 0D-3D mixed dimensions is
defined by the divergence of $a_\eff^{(\ell)}\to\infty$, which occurs
when
\begin{equation}\label{eq:resonance_0D}
 \frac1{\tilde a(\hat E_c)}\chi_\ell(r) = \frac{2\pi}{m_{AB}}
  \int\!d\r'\G_\ell(r;r')\big|_{k\to0}\chi_\ell(r')
\end{equation}
is satisfied.

\begin{figure*}[t]\hfill
 \includegraphics[width=0.46\textwidth,clip]{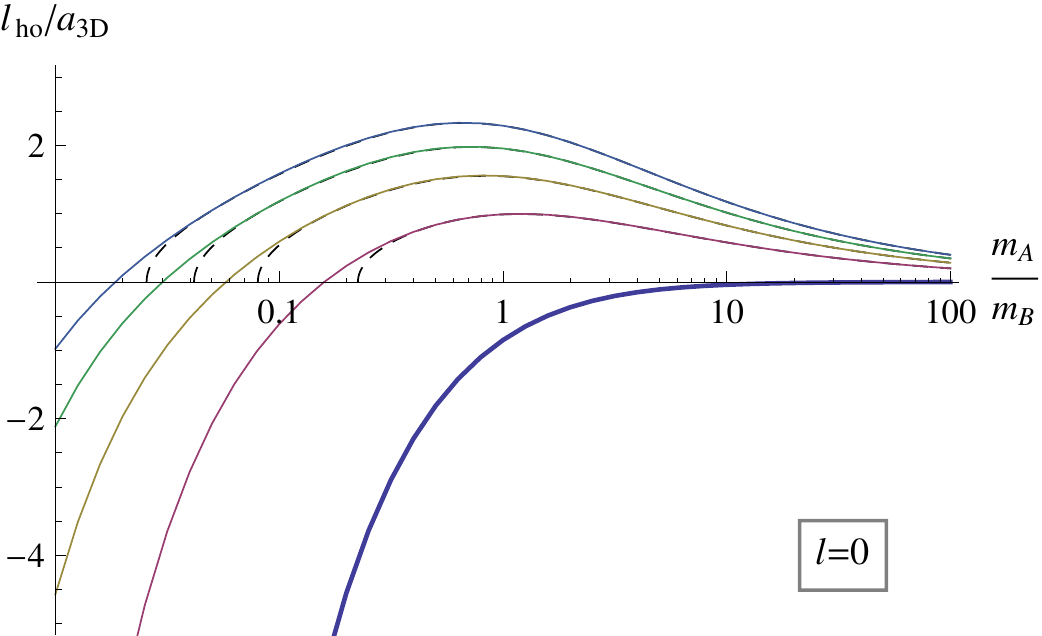}\hfill\hfill\hfill
 \includegraphics[width=0.46\textwidth,clip]{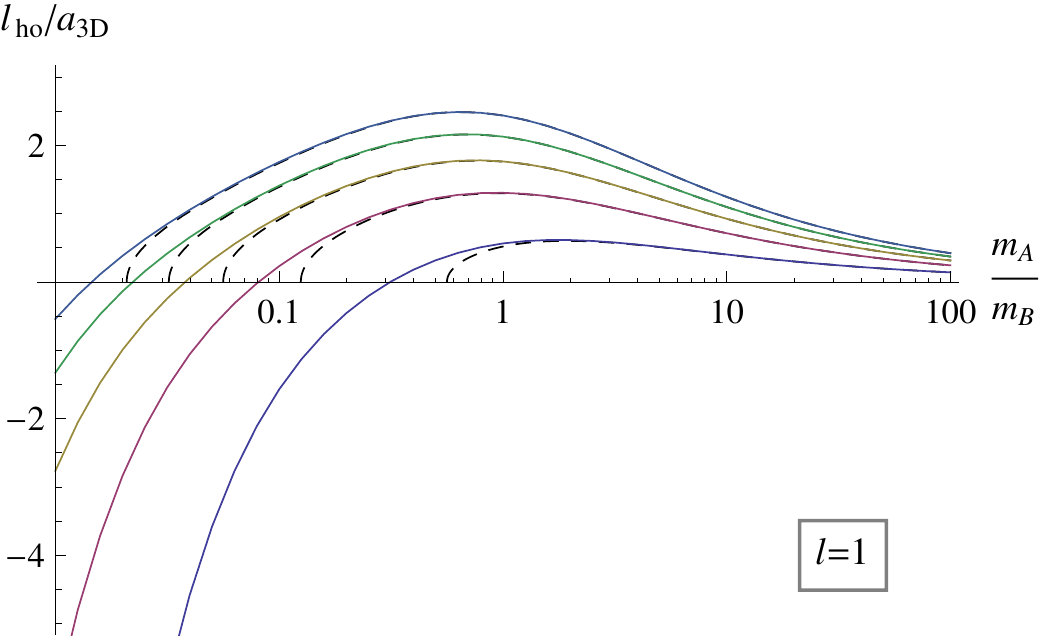}\hfill\vspace{4mm}\\\hfill
 \includegraphics[width=0.46\textwidth,clip]{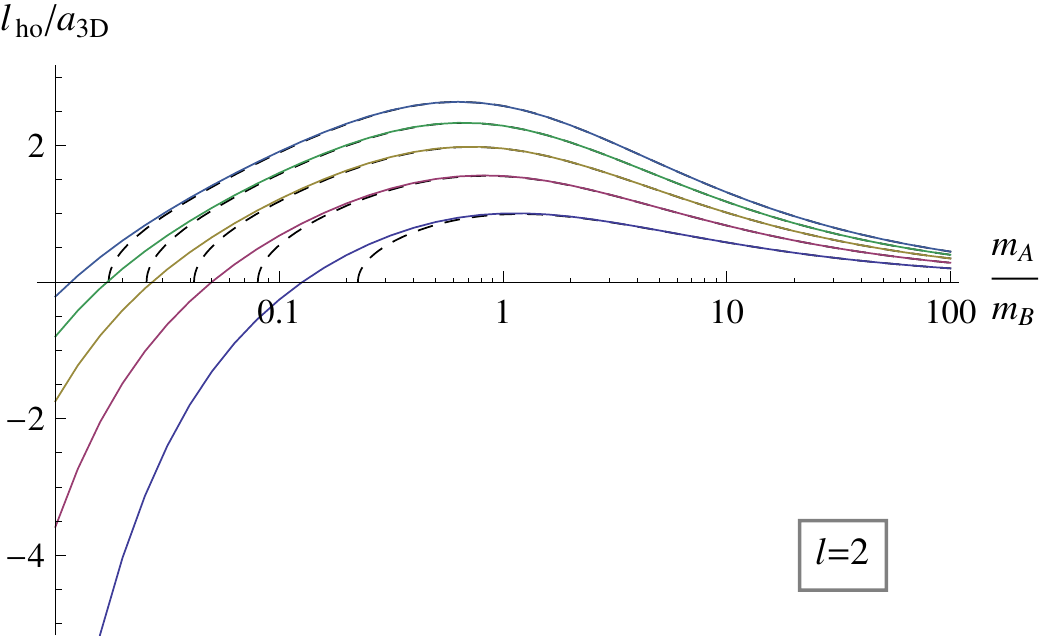}\hfill\hfill\hfill
 \includegraphics[width=0.46\textwidth,clip]{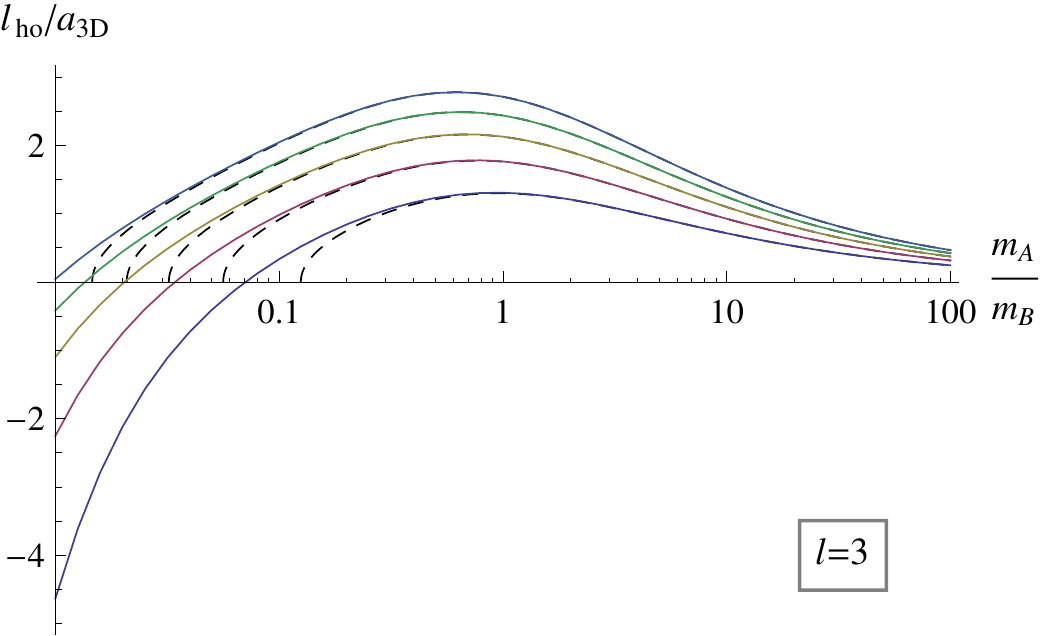}\hfill\hfill
 \caption{(Color online) 0D-3D mixture: Positions of the lowest five
 resonances in terms of $l_\mathrm{ho}/a_\mathrm{3D}$ for $\ell=0$
 (upper left), $\ell=1$ (upper right), $\ell=2$ (lower left), and
 $\ell=3$ (lower right) channels as functions of the mass ratio
 $m_A/m_B$.  The dashed curves are from the approximate formula
 $E_{AB}=\frac32\hbar\omega$ by using Eq.~(\ref{eq:approx_0D}) with
 $E_b=-\hbar^2/(2m_{AB}a_\mathrm{3D}^2)$.  \label{fig:resonance_0D}}
\end{figure*}

\subsection{Positions of resonances}
We now solve the integral equation (\ref{eq:resonance_0D}) numerically
to determine the positions of $\ell$th partial-wave resonances.
For the purpose of illustrating qualitative results, we shall set
$r_\mathrm{3D}=0$.  For quantitative predictions in a specific atomic
mixture, it is necessary but straightforward to include the effective
range correction~\cite{Lamporesi:2010}.  Some details of our method to
solve the integral equation are shown in Appendix~\ref{app:0D-3D}.

Figure~\ref{fig:resonance_0D} shows the positions of the lowest five
resonances in terms of $l_\mathrm{ho}/a_\mathrm{3D}$ for $\ell=0,1,2,3$
partial-wave channels as functions of the mass ratio $m_A/m_B$.  For
completeness, we have included the result for the $s$-wave ($\ell=0$)
resonance, which has been reported in Ref.~\cite{Massignan:2006}.  As we
have discussed in Sec.~\ref{sec:introduction}, there exists a series of
resonances in each partial-wave channel induced from the purely $s$-wave
interaction in a free space.  Indeed, the resonance positions are well
described by the approximate formula $E_{AB}=\frac32\hbar\omega$ by
using Eq.~(\ref{eq:approx_0D}) with
$E_b=-\hbar^2/(2m_{AB}a_\mathrm{3D}^2)$ in a wide range of the mass
ratio $m_A/m_B\gtrsim1$.  For such a mass ratio, as is evident from
Eq.~(\ref{eq:approx_0D}), $d$-wave ($f$-wave) resonances are nearly
degenerate with $s$-wave ($p$-wave) resonances, and thus it could be
difficult to distinguish them practically.  We also note that the
$\ell$th partial-wave resonance is ($2\ell+1$)-fold degenerate in a
spherically symmetric potential.  This degeneracy is lifted when the
confinement potential is deformed, and therefore the one resonance at a
given $l_\mathrm{ho}/a_\mathrm{3D}$ splits into more resonances.

\begin{figure*}[t]
 \begin{minipage}[t]{0.33\textwidth}
  \begin{center}\fbox{\scriptsize$m_A/m_B=6/40$}\end{center}
  \includegraphics[width=\textwidth,clip]{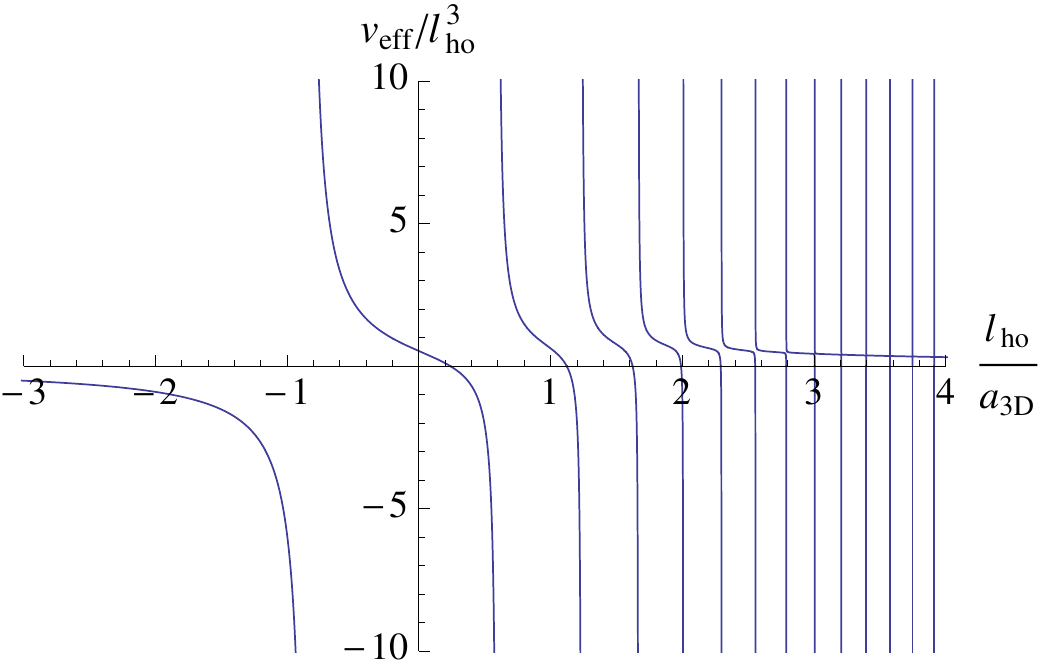}\vspace{4mm}
  \includegraphics[width=\textwidth,clip]{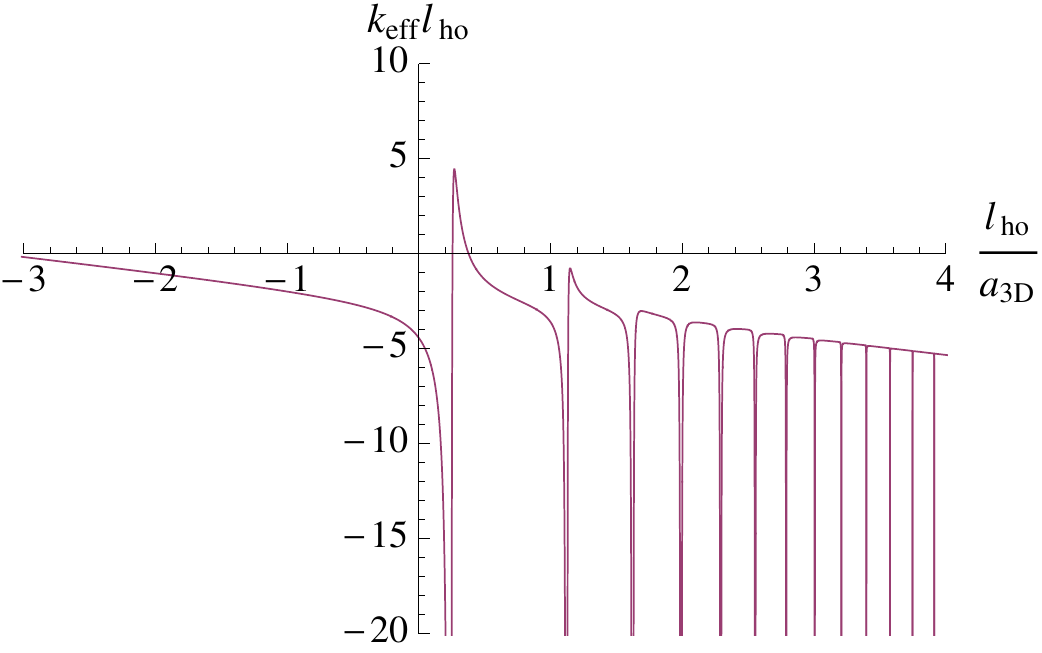}
 \end{minipage}\hfill
 \begin{minipage}[t]{0.33\textwidth}
  \begin{center}\fbox{\scriptsize$m_A/m_B=1$}\end{center}
  \includegraphics[width=\textwidth,clip]{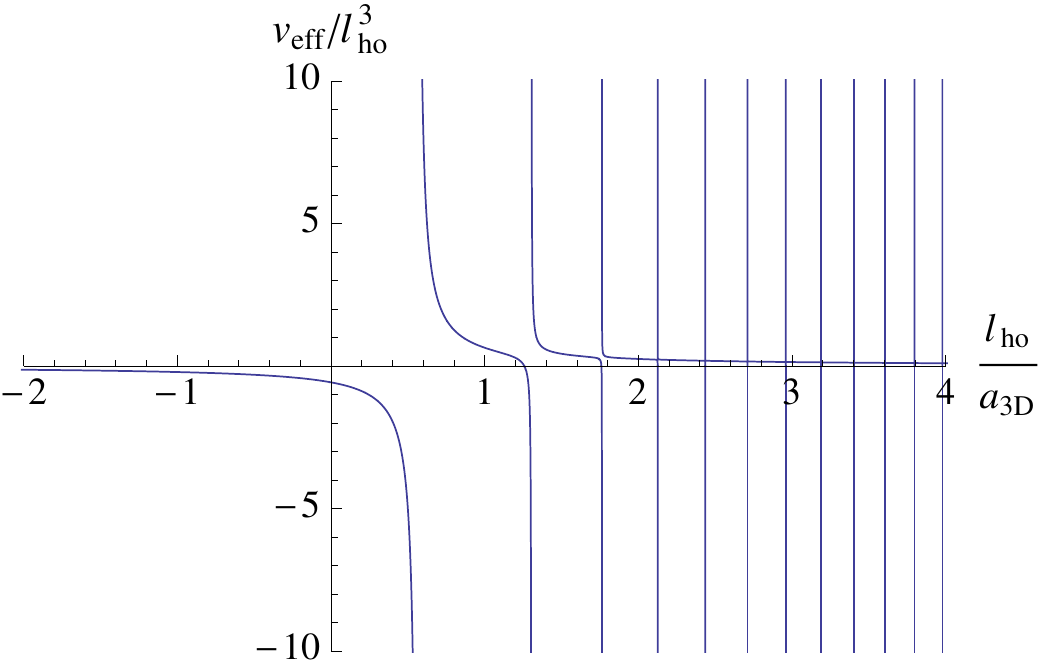}\vspace{4mm}
  \includegraphics[width=\textwidth,clip]{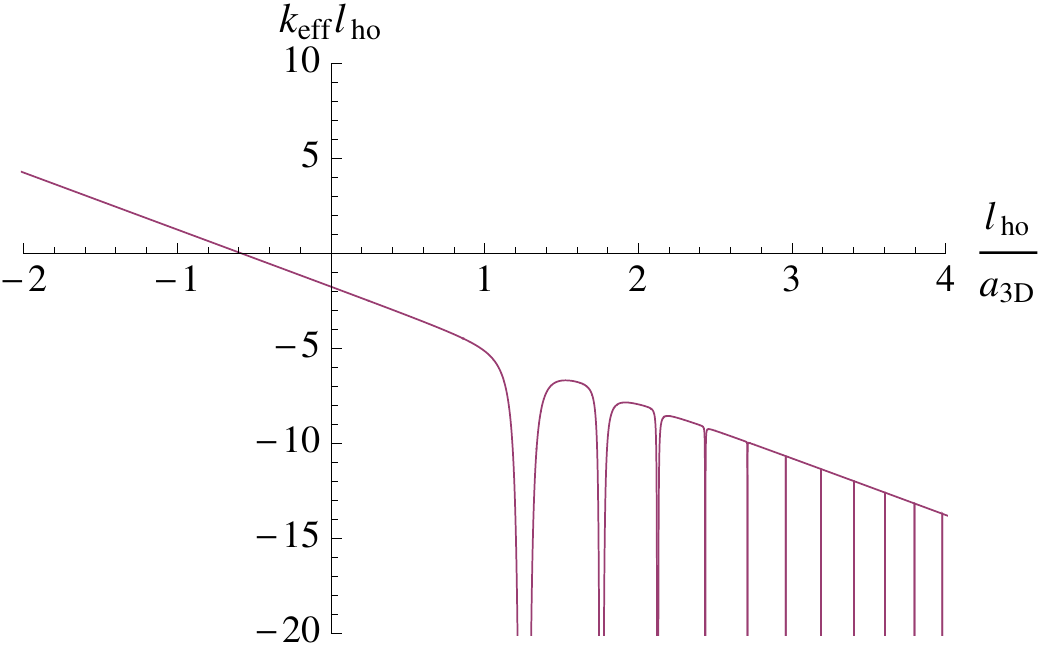}
 \end{minipage}\hfill
 \begin{minipage}[t]{0.33\textwidth}
  \begin{center}\fbox{\scriptsize$m_A/m_B=40/6$}\end{center}
  \includegraphics[width=\textwidth,clip]{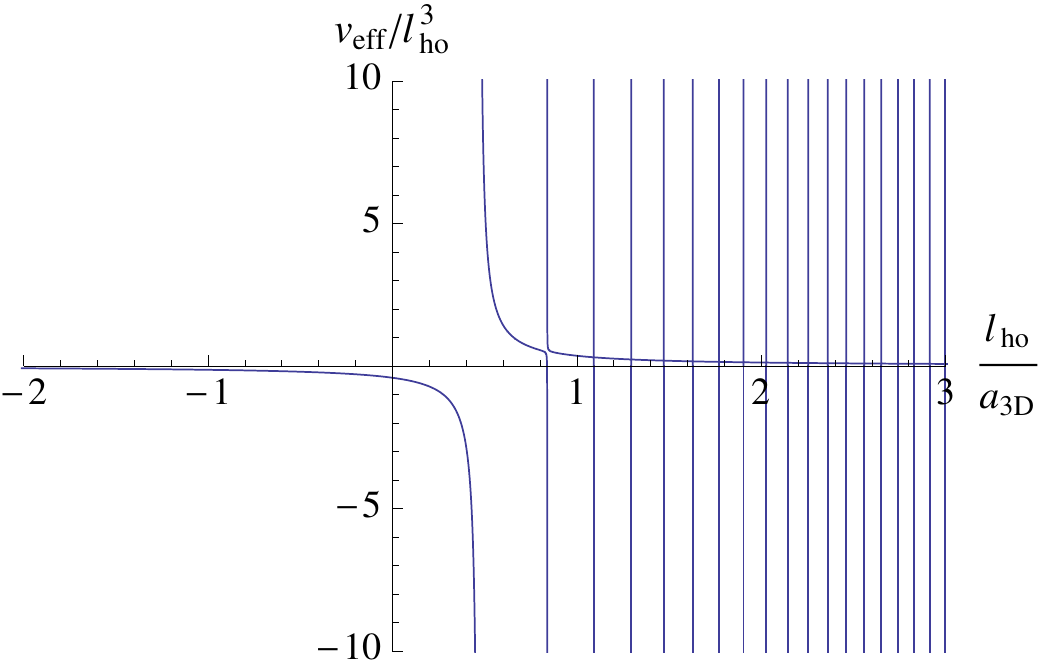}\vspace{4mm}
  \includegraphics[width=\textwidth,clip]{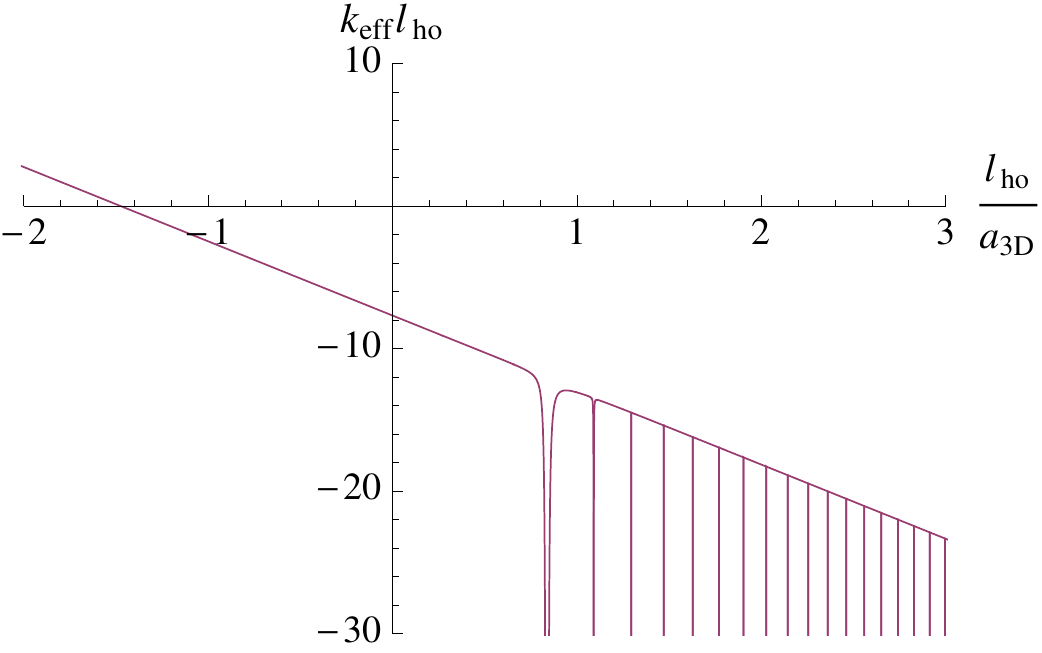}
 \end{minipage}
 \caption{(Color online) 0D-3D mixture: $p$-wave effective scattering
 volume $v_\eff/l_\mathrm{ho}^3$ (upper figures) and effective momentum
 $k_\eff l_\mathrm{ho}$ (lower figures) as functions of
 $l_\mathrm{ho}/a_\mathrm{3D}$ for mass ratios $m_A/m_B=6/40$ (left),
 $1$ (middle), and $40/6$ (right).  \label{fig:parameter_0D}}
\end{figure*}

\subsection{Confinement-induced molecules}
On the $a_\eff^{(\ell)}>0$ side of every resonance, a shallow $AB$
molecule is formed.  In the vicinity of the resonance
$a_\eff^{(\ell)}\gg l_\mathrm{ho}^{2\ell+1}$, its binding energy
$\varepsilon_{AB}\equiv E-\frac32\omega<0$ is determined by the pole of
the scattering amplitude [Eq.~(\ref{eq:expansion_0D})] with keeping the
two dominant terms at $k\to0$:
\begin{equation}\label{eq:binding_0D}
 \varepsilon_{AB} =
  \begin{cases}
   \displaystyle -\frac1{2m_Ba_\eff^{(\ell)2}} 
   &\ \text{for} \quad \ell=0, \medskip\\
   \displaystyle \phantom{\,}\frac1{m_Ba_\eff^{(\ell)}r_\eff^{(\ell)}}
   &\ \text{for} \quad \ell\geq1.
  \end{cases}
\end{equation}
Away from the resonance, these universal formulas are no longer valid.
The binding energy $\varepsilon_{AB}=-\kappa^2/(2m_B)$ has to be
determined by solving the integral equation
\begin{equation}
 \frac1{\tilde a(\hat E_c)}\chi_\ell^m(r) = \frac{2\pi}{m_{AB}}
  \int\!d\r'\G_\ell(r;r')\big|_{k\to i\kappa}\chi_\ell^m(r'),
\end{equation}
where $\chi_\ell^m$ is a component of the spherical harmonics expansion
of $\chi$:
\begin{equation}
 \chi(\r) = \sum_{\ell=0}^\infty\sum_{m=-\ell}^\ell\chi_\ell^m(r)\,Y_\ell^m(\hat\r).
\end{equation}
We note that the solution is independent of $m$, and therefore there are
$2\ell+1$ degenerate molecules for a spherically symmetric potential.
This degeneracy is lifted when the confinement potential is deformed.

We now derive the asymptotic form of the molecular wave function in the
vicinity of the $\ell$th partial-wave resonance where
$l_\mathrm{ho}\ll\kappa^{-1}\ll|\r_B|$ is satisfied.  From
Eqs.~(\ref{eq:solution_0D}) and (\ref{eq:G_0D}) with the replacement
$k\to i\kappa$, we find
\begin{equation}
 \begin{split}
  \psi(\r_A,\r_B) &\to -\sum_{m=-\ell}^\ell\frac1{2\ell+1}
  \frac{m_B}{m_{AB}}\frac{e^{-\kappa r_B}}{r_B^{\ell+1}}\phi_\0(r_A) \\
  &\quad \times Y_\ell^m(\hat\r_B)\int\!d\r'r'^\ell\phi_\0^*(r')\chi_\ell^m(r').
 \end{split}
\end{equation}
The angular parts of asymptotic wave functions of three degenerate
shallow $p$-wave molecules $\psi\sim Y_1^{0,\pm1}(\hat\r_B)$ are
illustrated in Fig.~\ref{fig:molecule}.

\subsection{Scattering parameters in the $p$-wave channel}
The effective scattering length in the $s$-wave channel
$a_\eff\equiv a_\eff^{(0)}$ has been computed in
Ref.~\cite{Massignan:2006}.  Here we focus on the $p$-wave ($\ell=1$)
channel and determine its two low-energy scattering parameters, namely,
the effective scattering volume $v_\eff\equiv a_\eff^{(1)}$ and the
effective momentum $k_\eff\equiv r_\eff^{(1)}$.  The effective
scattering volume can be computed by eliminating $C$ from
Eqs.~(\ref{eq:a_eff_0D-1}) and (\ref{eq:a_eff_0D-2}) and solving the
resulting integral equation numerically (see Appendix~\ref{app:0D-3D}
for details).  In Fig.~\ref{fig:parameter_0D}, $v_\eff/l_\mathrm{ho}^3$
for $r_\mathrm{3D}=0$ is plotted as a function of
$l_\mathrm{ho}/a_\mathrm{3D}$ for three mass ratios $m_A/m_B=6/40$, $1$,
and $40/6$.  We confirm the existence of a series of $p$-wave resonances
($v_\eff\to\infty$) induced from the purely $s$-wave interaction in a
free space, while they become narrower for larger
$l_\mathrm{ho}/a_\mathrm{3D}$.  We also find that the resonance is wider
when a lighter atom is confined in lower dimensions.

Similarly, the effective momentum can be computed from
Eq.~(\ref{eq:partial-wave_0D}), and $k_\eff l_\mathrm{ho}$ for
$r_\mathrm{3D}=0$ is plotted in Fig.~\ref{fig:parameter_0D} as a
function of $l_\mathrm{ho}/a_\mathrm{3D}$ for the same three mass
ratios.  In the vicinity of the $p$-wave resonance
$v_\eff\gg l_\mathrm{ho}^3$, $v_\eff$ and $k_\eff$ determine the binding
energy of three degenerate shallow $p$-wave molecules via the universal
formula (\ref{eq:binding_0D}).  Both $v_\eff$ and $k_\eff$ are important
to the low-energy effective theory of the $p$-wave resonance discussed
below.

\subsection{Low-energy effective theory}
The low-energy effective theory of the $p$-wave resonance in the 0D-3D
mixed dimensions is provided by the action
\begin{equation}\label{eq:action_0D}
 \begin{split}
  S &= \int\!dt\,\Psi_A^\+(t)\left(i\d_t\right)\Psi_A(t) \\
  &\quad + \int\!dtd\r\,\Psi_B^\+(t,\r)
  \left(i\d_t+\frac{\grad_{\!\r}^2}{2m_B}\right)\Psi_B(t,\r) \\
  &\quad + \int\!dt\,\Phi_j^\+(t)
  \left(i\d_t+\varepsilon_0\right)\Phi_j(t) \\
  &\quad + g_0\int\!dt
  \left[\Psi_A^\+(t)\nabla_{\!j}\Psi_B^\+(t,\0)\Phi_j(t)\right. \\
  &\qquad\qquad\quad
  \left.+\Phi_j^\+(t)\nabla_{\!j}\Psi_B(t,\0)\Psi_A(t)\right],
 \end{split}
\end{equation}
where the summation over $j=x,y,z$ is implicitly understood.  $\Psi_A$
and $\Psi_B$ fields represent the $A$ and $B$ atoms in 0D and 3D,
respectively.  The interaction between $A$ and $B$ atoms is described
through their coupling with three $p$-wave molecular fields $\Phi_j$.
$g_0$ is their coupling strength and $\varepsilon_0$ is the detuning
from the resonance.  These cutoff ($\Lambda$)-dependent bare parameters
can be related to the effective scattering volume $v_\eff$ and effective
momentum $k_\eff$ by matching the two-body scattering amplitude from the
action (\ref{eq:action_0D}) with that shown in
Eqs.~(\ref{eq:decomposition_0D}) and (\ref{eq:expansion_0D}).

The standard diagrammatic calculation leads to the following scattering
amplitude with collision energy $\varepsilon=k^2/(2m_B)$:
\begin{equation}
 \begin{split}
  i\mathcal{A}(k) = -\frac{i\k\cdot\k'}{\frac{k^2/(2m_B)+\varepsilon_0}{g_0^2}
  +\frac{m_B}{3\pi^2}\left(\frac{\Lambda^3}3+\Lambda k^2+\frac\pi2ik^3\right)}.
 \end{split}
\end{equation}
By defining
\begin{equation}
 \frac{\varepsilon_0}{g_0^2}+\frac{m_B}{3\pi^2}\frac{\Lambda^3}3
  \equiv \frac{m_B}{6\pi}\frac1{v_\eff}
\end{equation}
and
\begin{equation}
 \frac1{2m_Bg_0^2}+\frac{m_B}{3\pi^2}\Lambda
  \equiv -\frac{m_B}{6\pi}\frac{k_\eff}2,
\end{equation}
we reproduce the scattering amplitude (\ref{eq:expansion_0D}) in the
$p$-wave ($\ell=1$) channel up to a kinematical factor:
\begin{equation}
 \mathcal{A}(k) = -\frac{2\pi}{m_B}
  \frac{3\k\cdot\k'}{\frac1{v_\eff}-\frac{k_\eff}2k^2+ik^3}.
\end{equation}
This low-energy effective theory can be generalized easily to the case
with more than one lattice site where $B$ atoms are confined and could
be used to investigate the many-body physics across the $p$-wave
resonance as in the 3D case~\cite{Gurarie:2007}.  We note that an
effective field theory of the $p$-wave resonance in 3D has been
developed in connection with the $\alpha$-$n$
scattering~\cite{Bertulani:2002sz,Bedaque:2003wa}.

\subsection{Weak-coupling limit}
When $a_\mathrm{3D}<0$ and $|a_\mathrm{3D}|\ll l_\mathrm{ho}$, the
confinement-induced resonances can be understood in a different way, as
is discussed in Refs.~\cite{Massignan:2006,Vernier:2010}.  To the
leading order in the weak-coupling expansion
$a_\mathrm{3D}/l_\mathrm{ho}\to-0$, an $A$ atom occupies the ground
state in a 3D harmonic potential and creates a mean-field attractive
potential felt by a $B$ atom.  Therefore, the scattering of the $B$ atom
by the confined $A$ atom is described by
\begin{equation}\label{eq:weak-coupling_0D}
 \left[-\frac{\grad_{\!\r_B}^2}{2m_B}+\frac{2\pi a_\mathrm{3D}}{m_{AB}}
  |\phi_\0(r_B)|^2\right]\psi(\r_B) = \frac{k^2}{2m_B}\psi(\r_B),
\end{equation}
where
$|\phi_\0(r)|^2=e^{-r^2/l_\mathrm{ho}^2}/(\sqrt\pi\,l_\mathrm{ho})^3$.
This Schr\"odinger equation, which is valid in the weak-coupling limit,
is equivalent to the integral equation~(\ref{eq:chi_0D}), where only the
$\n=\0$ term is kept in the Green's function (\ref{eq:G_0D}) and
$\chi(\r)$ is identified as $a_\mathrm{3D}\phi_\0(r)\psi(\r)$.

By matching the solution of Eq.~(\ref{eq:weak-coupling_0D}) with the
asymptotic form (\ref{eq:asymptotic_0D}), we can determine the
scattering amplitude and low-energy scattering parameters in the
weak-coupling limit $|a_\mathrm{3D}|\ll l_\mathrm{ho}$.  In particular,
the resonance occurs when a new bound state appears.  This is possible
even in the weak-coupling limit because the attractive potential becomes
strong compared to the kinetic term by decreasing the mass ratio down to
$m_A/m_B\ll1$.  We find that the resonances are achieved at the critical
values of
$(m_B/m_{AB})(a_\mathrm{3D}/l_\mathrm{ho})=-1.19,-7.89,-20.2,\dots$ in
the $\ell=0$ channel~\cite{Massignan:2006,Vernier:2010},
$(m_B/m_{AB})(a_\mathrm{3D}/l_\mathrm{ho})=-5.36,-15.5,-31.2,\dots$ in
the $\ell=1$ channel,
$(m_B/m_{AB})(a_\mathrm{3D}/l_\mathrm{ho})=-11.9,-25.6,-44.6,\dots$ in
the $\ell=2$ channel, and
$(m_B/m_{AB})(a_\mathrm{3D}/l_\mathrm{ho})=-20.9,-37.9,-60.4,\dots$ in
the $\ell=3$ channel.

\section{1D-3D mixed dimensions \label{sec:1D-3D}}

\subsection{Scattering theory~\cite{normalization}}
The scattering of a quasi-1D $A$ atom with a $B$ atom in 3D is described
by a Schr\"odinger equation,
\begin{equation}\label{eq:schrodinger_1D}
 \begin{split}
  & \left(-\frac{\grad_{\!\bm\rho_A}^2}{2m_A}+\frac12m_A\omega^2\bm\rho_A^2
  -\frac{\grad_{\!\bm\rho_B}^2}{2m_B}-\frac{\nabla_{\!z_{AB}}^2}{2m_{AB}}\right) \\
  &\times \psi(\bm\rho_A,\bm\rho_B,z_{AB}) = E\,\psi(\bm\rho_A,\bm\rho_B,z_{AB})
 \end{split}
\end{equation}
for $\sqrt{(\bm\rho_A-\bm\rho_B)^2+z_{AB}^2}>0$, where
$\bm\rho\equiv(x,y)$, $z_{AB}\equiv z_A-z_B$, and the center-of-mass
motion in the $z$ direction is eliminated.  The short-range interspecies
interaction is implemented by the generalized Bethe-Peierls boundary
condition~\cite{Petrov:2004,Levinsen:2009mn}:
\begin{equation}\label{eq:short-range_1D}
 \begin{split}
  & \psi(\bm\rho_A,\bm\rho_B,z_{AB})\big|_{\bm\rho_A,\bm\rho_B\to\bm\rho;z_{AB}\to0} \\
  &\to \left[\frac1{\tilde a(\hat E_c)}
  -\frac1{\sqrt{(\bm\rho_A-\bm\rho_B)^2+z_{AB}^2}}\right]\chi(\bm\rho).
 \end{split}
\end{equation}
The collision energy operator $\hat E_c$ in Eq.~(\ref{eq:generalized_a})
in the present case is given by
\begin{equation}
 \hat E_c = E
  - \left(-\frac{\grad_{\!\bm\rho}^2}{2M}+\frac12m_A\omega^2\bm\rho^2\right).
\end{equation}

The solution to the Schr\"odinger equation (\ref{eq:schrodinger_1D}) can
be written as
\begin{equation}\label{eq:solution_1D}
 \begin{split}
  & \psi(\bm\rho_A,\bm\rho_B,z_{AB}) = \psi_0(\bm\rho_A,\bm\rho_B,z_{AB}) \\
  &\quad + \frac{2\pi}{m_{AB}}\int\!d\bm\rho'
  G_E(\bm\rho_A,\bm\rho_B,z_{AB};\bm\rho',\bm\rho',0)\chi(\bm\rho'),
 \end{split}
\end{equation}
where $\psi_0$ is a solution in the noninteracting limit and $G_E$ is the
retarded Green's function for the noninteracting Hamiltonian:
\begin{equation}\label{eq:G_1D}
 \begin{split}
  & G_E(\bm\rho_A,\bm\rho_B,z_{AB};\bm\rho'_A,\bm\rho'_B,z'_{AB}) \\
  &\equiv \<\bm\rho_A,\bm\rho_B,z_{AB}|\frac1{E-H_0+i0^+}|\bm\rho'_A,\bm\rho'_B,z'_{AB}\> \\
  &= -\frac{\sqrt{m_Bm_{AB}}}{2\pi}\sum_\n
  \phi_\n(\bm\rho_A)\phi_\n^*(\bm\rho'_A) \\
  &\quad \times \frac{e^{-\sqrt{2m_B}\sqrt{(n_x+n_y+1)\omega-E-i0^+}
  |\tilde\r_B-\tilde\r_B'|}}{|\tilde\r_B-\tilde\r_B'|}.
 \end{split}
\end{equation}
Here $\phi_\n$ with quantum numbers $\n=(n_x,n_y)$ is the normalized wave
function of an $A$ atom in the 2D harmonic potential, and 
$\tilde\r_B\equiv\left(\bm\rho_B,-\sqrt\frac{m_{AB}}{m_B}\,z_{AB}\right)$ are
coordinates of the $B$ atom relative to the confined $A$ atom.  The
anisotropic factor is such because separations in the $x$ and $y$
directions are associated with $m_B$ while a separation in the $z$
direction is associated with $m_{AB}$ [see the last two terms in
Eq.~(\ref{eq:schrodinger_1D})].

We now consider the low-energy scattering in which
\begin{equation}
 E-\omega \equiv \frac{k^2}{2m_B} \ll \omega
\end{equation}
is satisfied, and then $\psi_0$ becomes
\begin{equation}
 \psi_0(\bm\rho_A,\bm\rho_B,z_{AB}) = Ce^{i\k\cdot\tilde\r_B}\phi_\0(\rho_A),
\end{equation}
which represents the $A$ atom in the ground state of the 2D harmonic
potential and the plane wave of $B$ atom with the wave vector $\k$.  The
asymptotic form of the wave function at a large distance
$|\tilde\r_B|\gg l_\mathrm{ho}$ is given by
\begin{equation}\label{eq:asymptotic_1D}
 \psi(\bm\rho_A,\bm\rho_B,z_{AB}) \to C\left[e^{i\k\cdot\tilde\r_B}
  +\frac{e^{ik\tilde r_B}}{\tilde r_B}f(\k_\perp,\k'_\perp)\right]\phi_\0(\rho_A),
\end{equation}
where $f(\k_\perp,\k'_\perp)$ with $\k'\equiv k\bm\hat{\tilde\r}_B$
defines the two-body scattering amplitude in the 1D-3D mixed dimensions:
\begin{equation}\label{eq:f_1D}
 f(\k_\perp,\k'_\perp) \equiv -\frac1{C}\sqrt\frac{m_B}{m_{AB}}
  \int\!d\bm\rho'e^{-i\k'_\perp\cdot\bm\rho'}\phi_\0^*(\rho')\chi(\bm\rho').
\end{equation}
We note that $\chi$ has an implicit $\k_\perp\equiv(k_x,k_y)$
dependence and both $\chi$ and $f$ depend on $k$ through the Green's
function (\ref{eq:G_1D}).

The unknown function $\chi$ can be determined by substituting the
solution (\ref{eq:solution_1D}) into the Bethe-Peierls boundary
condition (\ref{eq:short-range_1D}).  Defining the regular part of the
Green's function $\G$ by
\begin{equation}\label{eq:regular_1D}
 \begin{split}
  & G_E(\bm\rho_A,\bm\rho_B,z_{AB};\bm\rho',\bm\rho',0)
  \big|_{\bm\rho_A,\bm\rho_B\to\bm\rho;z_{AB}\to0} \\
  &\equiv -\frac{m_{AB}}{2\pi\sqrt{(\bm\rho_A-\bm\rho_B)^2+z_{AB}^2}}
  \delta(\bm\rho-\bm\rho') + \G(\bm\rho;\bm\rho'),
 \end{split}
\end{equation}
we obtain
\begin{equation}\label{eq:chi_1D}
 \frac1{\tilde a(\hat E_c)}\chi(\bm\rho) = C e^{i\k_\perp\cdot\bm\rho}\phi_\0(\rho)
  + \frac{2\pi}{m_{AB}}\int\!d\bm\rho'\G(\bm\rho;\bm\rho')\chi(\bm\rho').
\end{equation}
This integral equation determines $\chi/C$, which in turn provides $f$
from Eq.~(\ref{eq:f_1D}).

Because the system has a rotational symmetry in the $xy$ plane, $f$,
$\chi$, and $\G$ can be decomposed into their partial-wave components:
\begin{align}\label{eq:decomposition_1D}
 f(\k_\perp,\k'_\perp) &= \sum_{m=-\infty}^\infty
 f_m(k_\perp,k'_\perp)\,e^{im\arccos\hat\k_\perp\cdot\hat\k'_\perp}, \\
 \chi(\bm\rho) &= \sum_{m=-\infty}^\infty
 \chi_m(\rho)\,e^{im\arccos\hat{\bm\rho}\cdot\hat\k_\perp}, \\
 \G(\bm\rho;\bm\rho') &= \sum_{m=-\infty}^\infty
 \G_m(\rho;\rho')\,e^{im\arccos\hat{\bm\rho}\cdot\hat{\bm\rho}'}.
\end{align}
Because $m$ and $-m$ are degenerate, we assume $m\geq0$ in this
section without loss of generality.  Equations~(\ref{eq:f_1D}) and
(\ref{eq:chi_1D}) lead to the $m$th partial-wave scattering amplitude
given by
\begin{equation}\label{eq:partial-wave_1D}
 f_m(k_\perp,k'_\perp) = -\frac1{C}\sqrt\frac{m_B}{m_{AB}}
  \int\!d\bm\rho'J_m(k_\perp\rho')\phi_\0^*(\rho')\chi_m(\rho')
\end{equation}
with
\begin{equation}
 \begin{split}
  \frac1{\tilde a(\hat E_c)}\chi_m(\rho) &= CJ_m(k_\perp\rho)\phi_\0(\rho) \\
  &\quad + \frac{2\pi}{m_{AB}}\int\!d\bm\rho'\G_m(\rho;\rho')\chi_m(\rho').
 \end{split}
\end{equation}
From the explicit calculation that uses the Green's function in
Eq.~(\ref{eq:G_1D}), we can show that $f_m$ has the following low-energy
expansion:
\begin{equation}\label{eq:expansion_1D}
 \begin{split}
  &\lim_{k\to0}f_m(k) \\
  &= -\frac{\frac{(2m+1)!}{(2^mm!)^2}\,k^m_\perp k'^m_\perp}
  {\frac1{a_\eff^{(m)}}-\frac12r_\eff^{(m)}k^2+O(k^4)+ik^{2m+1}\left[1+O(k^2)\right]} \\
  &\quad + O(k^{m+2}_\perp k'^m_\perp,k^m_\perp k'^{m+2}_\perp),
 \end{split}
\end{equation}
where $a_\eff^{(m)}$ and $r_\eff^{(m)}$ are effective scattering
``length'' and ``range'' parameters in the $m$th partial-wave channel.
Note that $a_\eff^{(m)}$ has the dimension of $(\mathrm{length})^{2m+1}$
and $r_\eff^{(m)}$ has $(\mathrm{length})^{1-2m}$.  This unusual form of
the low-energy expansion is owing to the lack of full 3D rotational
symmetries.

\begin{figure*}[t]\hfill
 \includegraphics[width=0.46\textwidth,clip]{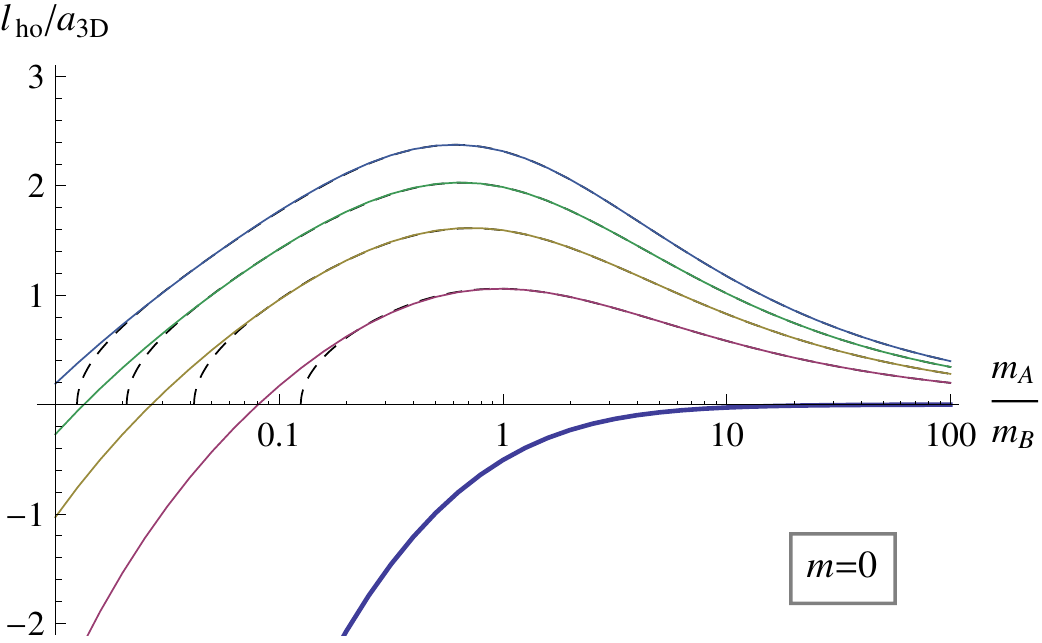}\hfill\hfill\hfill
 \includegraphics[width=0.46\textwidth,clip]{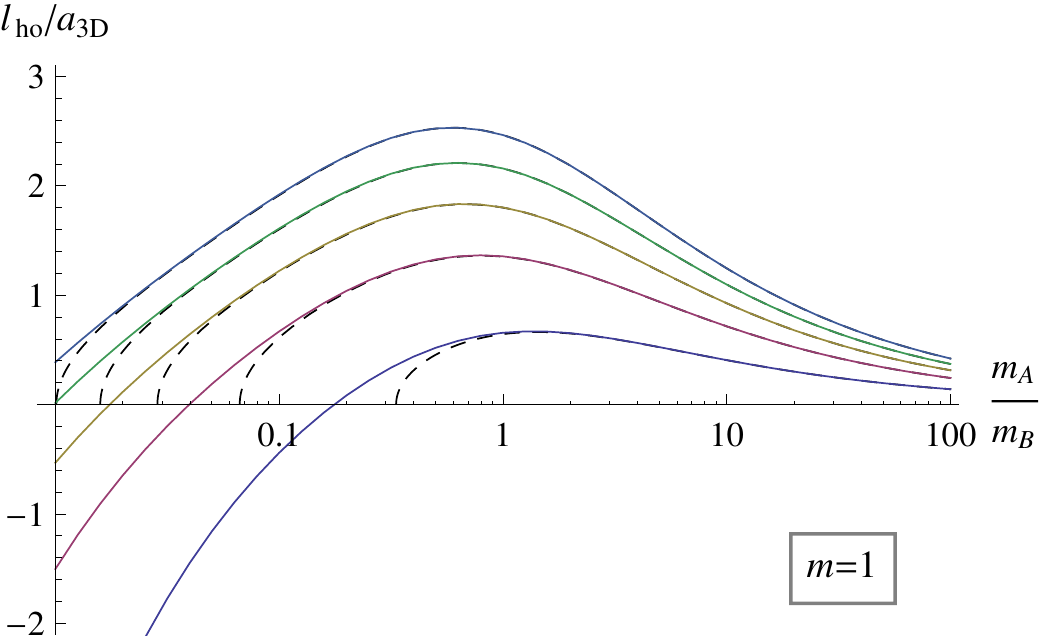}\hfill\vspace{4mm}\\\hfill
 \includegraphics[width=0.46\textwidth,clip]{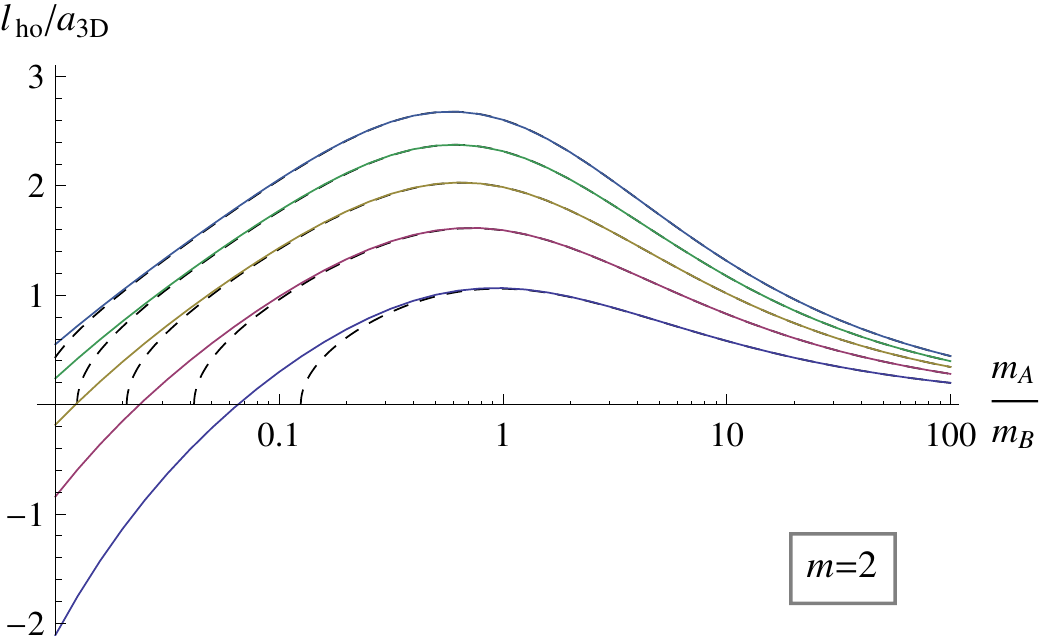}\hfill\hfill\hfill
 \includegraphics[width=0.46\textwidth,clip]{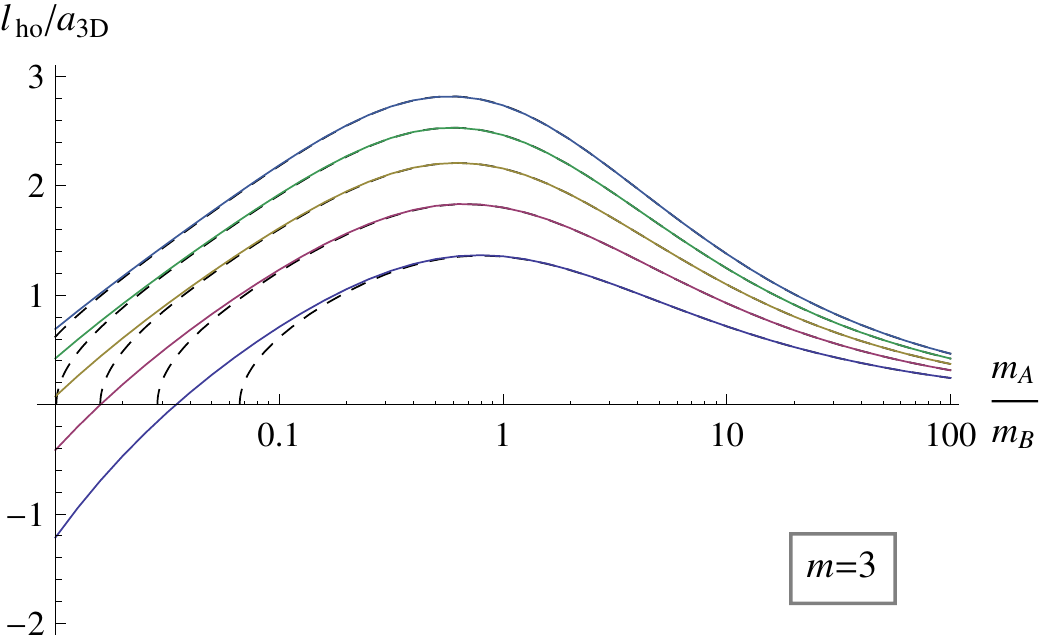}\hfill\hfill
 \caption{(Color online) 1D-3D mixture: Positions of the lowest five
 resonances in terms of $l_\mathrm{ho}/a_\mathrm{3D}$ for $m=0$ (upper
 left), $m=1$ (upper right), $m=2$ (lower left), and $m=3$ (lower right)
 channels as functions of the mass ratio $m_A/m_B$.  The dashed curves
 are from the approximate formula $E_{AB}=\hbar\omega$ by using
 Eq.~(\ref{eq:approx_1D}) with $E_b=-\hbar^2/(2m_{AB}a_\mathrm{3D}^2)$.
 \label{fig:resonance_1D}}
\end{figure*}

Substituting the expansion of $f_m$ into Eq.~(\ref{eq:partial-wave_1D}),
we can determine the low-energy scattering parameters.  In particular,
the effective scattering length $a_\eff^{(m)}$ is given by
\begin{equation}\label{eq:a_eff_1D-1}
 \frac{(2m+1)!}{(2^mm!)^2}a_\eff^{(m)} = \frac1{C}\sqrt\frac{m_B}{m_{AB}}
  \frac1{2^mm!}\int\!d\bm\rho'\rho'^m\phi_\0^*(\rho')\chi_m(\rho')
\end{equation}
with
\begin{equation}\label{eq:a_eff_1D-2}
 \begin{split}
  \frac1{\tilde a(\hat E_c)}\chi_m(\rho) &= C\frac1{2^mm!}\rho^m\phi_\0(\rho) \\
  &\quad + \frac{2\pi}{m_{AB}}\int\!d\bm\rho'\G_m(\rho;\rho')\big|_{k\to0}\chi_m(\rho').
 \end{split}
\end{equation}
The $m$th partial-wave resonance in the 1D-3D mixed dimensions is
defined by the divergence of $a_\eff^{(m)}\to\infty$, which occurs when
\begin{equation}\label{eq:resonance_1D}
 \frac1{\tilde a(\hat E_c)}\chi_m(\rho) = \frac{2\pi}{m_{AB}}
  \int\!d\bm\rho'\G_m(\rho;\rho')\big|_{k\to0}\chi_m(\rho')
\end{equation}
is satisfied.

\subsection{Positions of resonances}
We now solve the integral equation (\ref{eq:resonance_1D}) numerically
to determine the positions of $|m|$th partial-wave resonances.  For the
purpose of illustrating qualitative results, we shall set
$r_\mathrm{3D}=0$.  For quantitative predictions in a specific atomic
mixture, it is necessary but straightforward to include the effective
range correction~\cite{Lamporesi:2010}.  Some details of our method to
solve the integral equation are shown in Appendix~\ref{app:1D-3D}.

Figure~\ref{fig:resonance_1D} shows the positions of the lowest five
resonances in terms of $l_\mathrm{ho}/a_\mathrm{3D}$ for $m=0,1,2,3$
partial-wave channels as functions of the mass ratio $m_A/m_B$.  For
completeness, we have included the result for the $s$-wave ($m=0$)
resonance, which has been reported in Ref.~\cite{Nishida:2008kr}.  As we
have discussed in Sec.~\ref{sec:introduction}, there exists a series of
resonances in each partial-wave channel induced from the purely $s$-wave
interaction in a free space.  Indeed, the resonance positions are well
described by the approximate formula $E_{AB}=\hbar\omega$ by using
Eq.~(\ref{eq:approx_1D}) with $E_b=-\hbar^2/(2m_{AB}a_\mathrm{3D}^2)$ in
a wide range of the mass ratio $m_A/m_B\gtrsim1$.  For such a mass
ratio, as is evident from Eq.~(\ref{eq:approx_1D}), $d$-wave ($f$-wave)
resonances are nearly degenerate with $s$-wave ($p$-wave) resonances,
and thus it could be difficult to distinguish them practically.  We also
note that the $|m|$th partial-wave resonance for $|m|\geq1$ is twofold
degenerate in a circularly symmetric potential.  This degeneracy is
lifted when the confinement potential is deformed, and therefore the one
resonance at a given $l_\mathrm{ho}/a_\mathrm{3D}$ splits into two
resonances.

\begin{figure*}[t]
 \begin{minipage}[t]{0.33\textwidth}
  \begin{center}\fbox{\scriptsize$m_A/m_B=6/40$}\end{center}
  \includegraphics[width=\textwidth,clip]{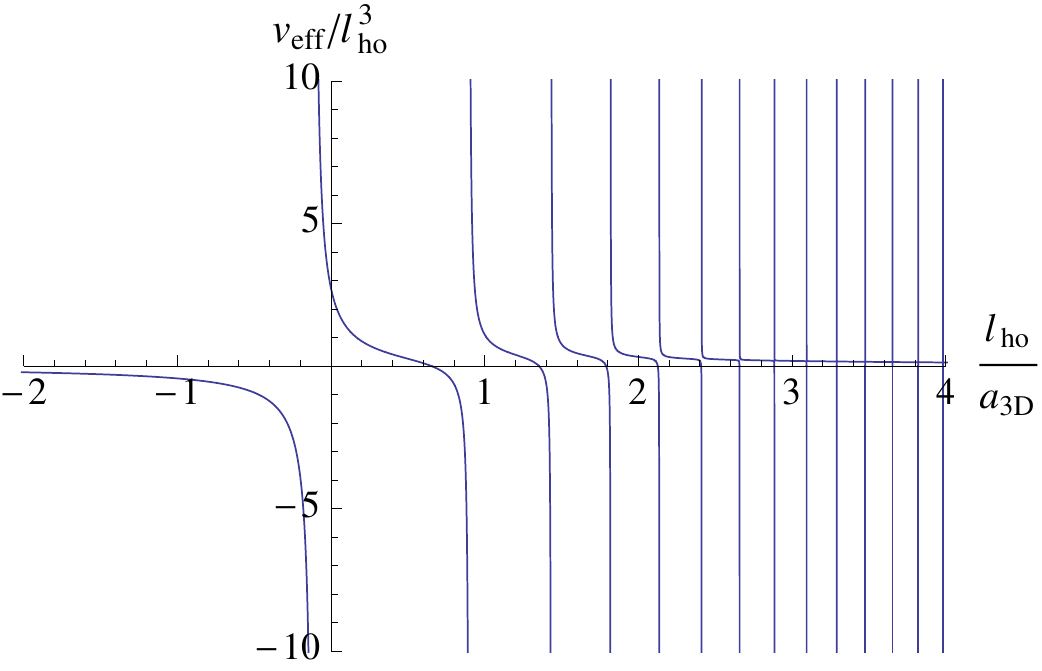}\vspace{4mm}
  \includegraphics[width=\textwidth,clip]{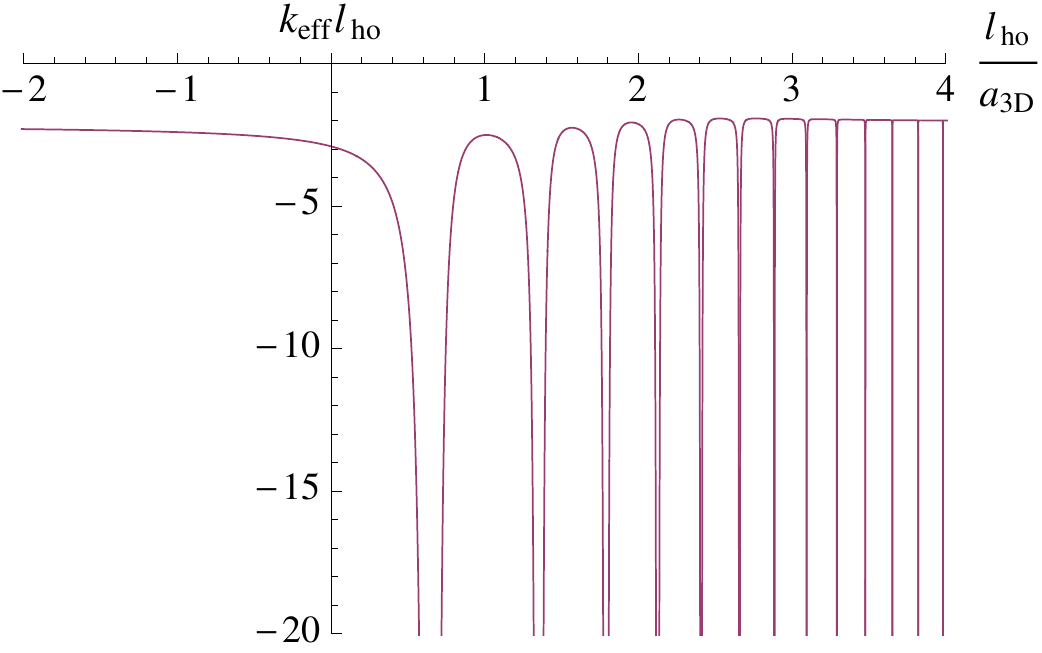}
 \end{minipage}\hfill
 \begin{minipage}[t]{0.33\textwidth}
  \begin{center}\fbox{\scriptsize$m_A/m_B=1$}\end{center}
  \includegraphics[width=\textwidth,clip]{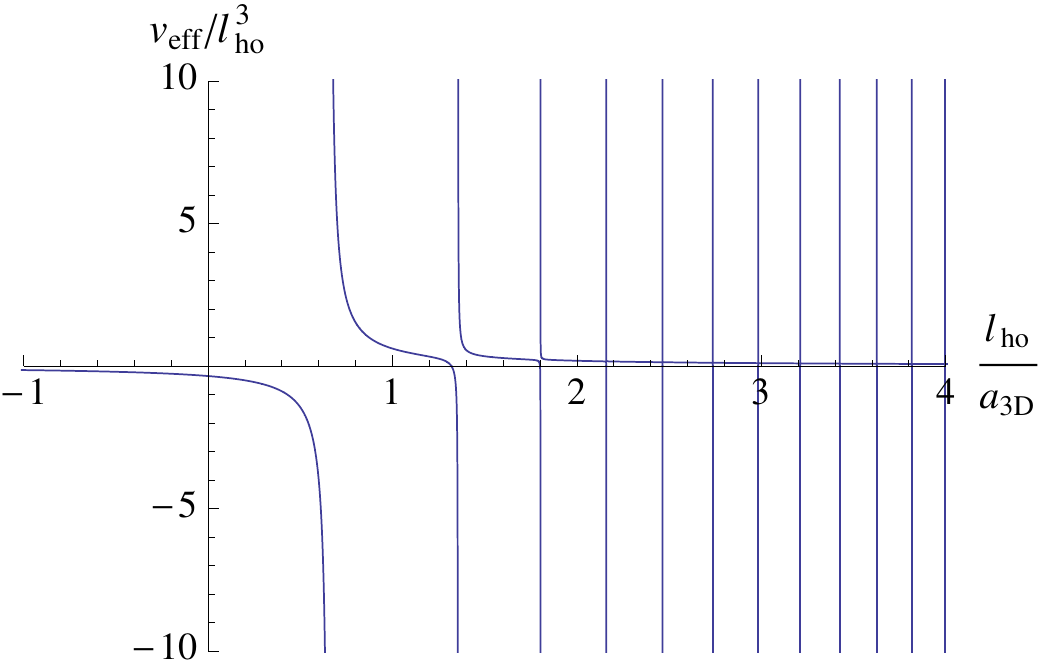}\vspace{4mm}
  \includegraphics[width=\textwidth,clip]{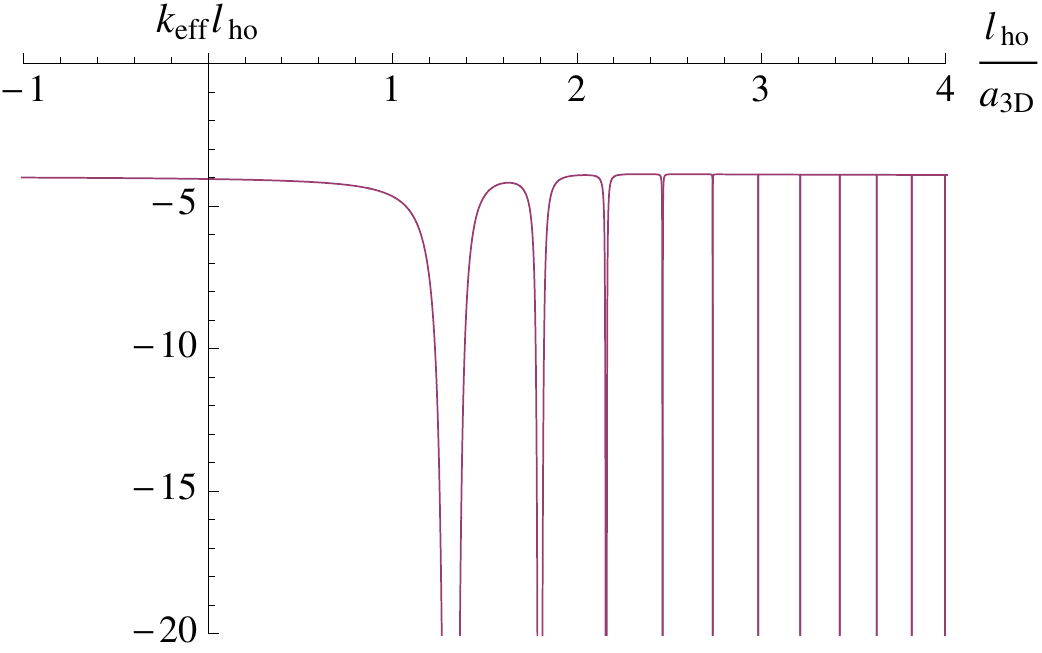}
 \end{minipage}\hfill
 \begin{minipage}[t]{0.33\textwidth}
  \begin{center}\fbox{\scriptsize$m_A/m_B=40/6$}\end{center}
  \includegraphics[width=\textwidth,clip]{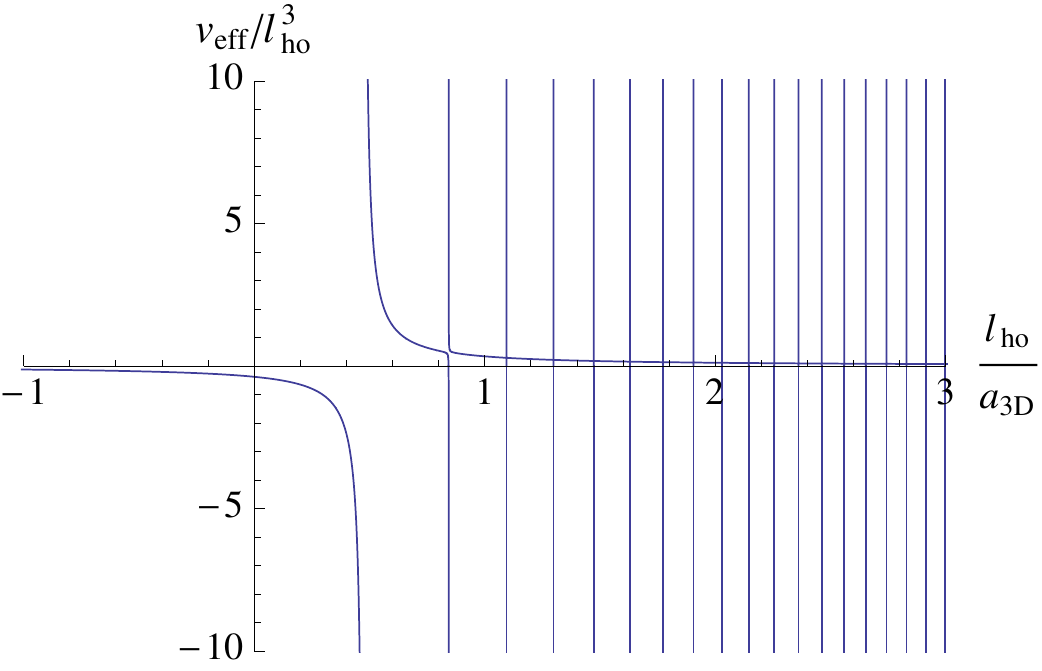}\vspace{4mm}
  \includegraphics[width=\textwidth,clip]{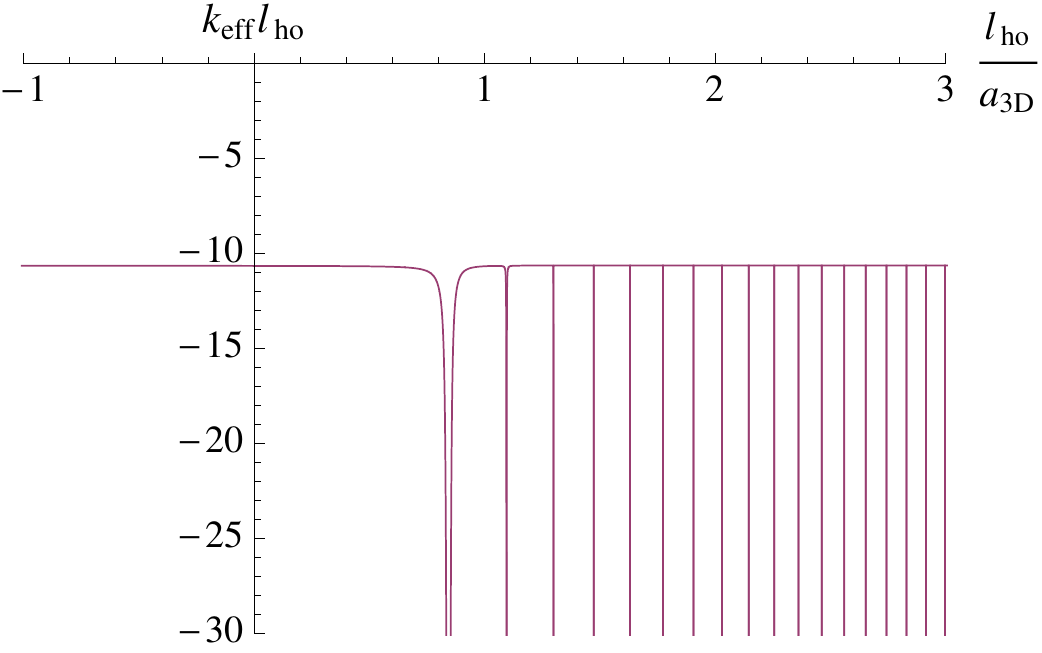}
 \end{minipage}
 \caption{(Color online) 1D-3D mixture: $p$-wave effective scattering
 volume $v_\eff/l_\mathrm{ho}^3$ (upper figures) and effective momentum
 $k_\eff l_\mathrm{ho}$ (lower figures) as functions of
 $l_\mathrm{ho}/a_\mathrm{3D}$ for mass ratios $m_A/m_B=6/40$ (left),
 $1$ (middle), and $40/6$ (right).  \label{fig:parameter_1D}}
\end{figure*}

\subsection{Confinement-induced molecules}
On the $a_\eff^{(m)}>0$ side of every resonance, a shallow $AB$
molecule is formed.  In the vicinity of the resonance
$a_\eff^{(m)}\gg l_\mathrm{ho}^{2|m|+1}$, its binding energy
$\varepsilon_{AB}\equiv E-\omega<0$ is determined by the pole of the
scattering amplitude [Eq.~(\ref{eq:expansion_1D})] with keeping the two
dominant terms at $k\to0$:
\begin{equation}\label{eq:binding_1D}
 \varepsilon_{AB} =
  \begin{cases}
   \displaystyle -\frac1{2m_Ba_\eff^{(m)2}} 
   &\ \text{for} \quad m=0, \medskip\\
   \displaystyle \phantom{\,}\frac1{m_Ba_\eff^{(m)}r_\eff^{(m)}}
   &\ \text{for} \quad |m|\geq1.
  \end{cases}
\end{equation}
Away from the resonance, these universal formulas are no longer valid.
The binding energy $\varepsilon_{AB}=-\kappa^2/(2m_B)$ has to be
determined by solving the integral equation
\begin{equation}
 \frac1{\tilde a(\hat E_c)}\chi_m(\rho) = \frac{2\pi}{m_{AB}}
  \int\!d\bm\rho'\G_m(\rho;\rho')\big|_{k\to i\kappa}\chi_m(\rho').
\end{equation}
We note that the solution for $|m|\geq1$ is independent of the sign of
$m$, and therefore there are two degenerate molecules for a circularly
symmetric potential.  This degeneracy is lifted when the confinement
potential is deformed.

We now derive the asymptotic form of the molecular wave function in the
vicinity of the $|m|$th partial-wave resonance where
$l_\mathrm{ho}\ll\kappa^{-1}\ll|\tilde\r_B|$ is satisfied.  From
Eqs.~(\ref{eq:solution_1D}) and (\ref{eq:G_1D}) with the replacement
$k\to i\kappa$, we find
\begin{equation}
 \begin{split}
  & \psi(\bm\rho_A,\bm\rho_B,z_{AB}) \\
  &\to -\sum_{m=\pm|m|}\sum_{\ell=|m|}^\infty
  \sqrt{\frac{4\pi}{2\ell+1}\frac{(\ell-m)!}{(\ell+m)!}}P_\ell^m(0)  
  \sqrt\frac{m_B}{m_{AB}} \\
  &\quad \times \frac{e^{-\kappa\tilde r_B}}{\tilde r_B^{\ell+1}}
  \phi_\0(\rho_A)Y_\ell^m(\hat{\tilde\r}_B)
  \int\!d\bm\rho'\rho'^\ell\phi_\0^*(\rho')\chi_m(\rho').
 \end{split}
\end{equation}
One can see that different spherical harmonics $\sim Y_\ell^m$ with the
same magnetic quantum number $m$ contribute owing to the lack of full 3D
rotational symmetries.  The $s$-wave nature of the free-space
interaction ensures that only $\ell=|m|,|m|+2,|m|+4,\dots$ contribute
so that the wave function is an even function of $z_{AB}$.  The
asymptotic behavior at a large separation $|\tilde\r_B|\to\infty$ is
dominated by the $\ell=|m|$ component.  The angular parts of asymptotic
wave functions of two degenerate shallow $p$-wave molecules
$\psi\sim Y_1^{\pm1}(\hat{\tilde\r}_B)$ are illustrated in
Fig.~\ref{fig:molecule}.

\subsection{Scattering parameters in the $p$-wave channel}
The effective scattering length in the $s$-wave channel
$a_\eff\equiv a_\eff^{(0)}$ has been computed in
Ref.~\cite{Nishida:2008kr}.  Here we focus on the $p$-wave ($|m|=1$)
channel and determine its two low-energy scattering parameters, namely,
the effective scattering volume $v_\eff\equiv a_\eff^{(1)}$ and the
effective momentum $k_\eff\equiv r_\eff^{(1)}$.  The effective
scattering volume can be computed by eliminating $C$ from
Eqs.~(\ref{eq:a_eff_1D-1}) and (\ref{eq:a_eff_1D-2}) and solving the
resulting integral equation numerically (see Appendix~\ref{app:1D-3D}
for details).  In Fig.~\ref{fig:parameter_1D}, $v_\eff/l_\mathrm{ho}^3$
for $r_\mathrm{3D}=0$ is plotted as a function of
$l_\mathrm{ho}/a_\mathrm{3D}$ for three mass ratios $m_A/m_B=6/40$, $1$,
and $40/6$.  We confirm the existence of a series of $p$-wave resonances
($v_\eff\to\infty$) induced from the purely $s$-wave interaction in a
free space, while they become narrower for larger
$l_\mathrm{ho}/a_\mathrm{3D}$.  We also find that the resonance is wider
when a lighter atom is confined in lower dimensions.

Similarly, the effective momentum can be computed from
Eq.~(\ref{eq:partial-wave_1D}), and $k_\eff l_\mathrm{ho}$ for
$r_\mathrm{3D}=0$ is plotted in Fig.~\ref{fig:parameter_1D} as a
function of $l_\mathrm{ho}/a_\mathrm{3D}$ for the same three mass
ratios.  In the vicinity of the $p$-wave resonance
$v_\eff\gg l_\mathrm{ho}^3$, $v_\eff$ and $k_\eff$ determine the binding
energy of two degenerate shallow $p$-wave molecules via the universal
formula (\ref{eq:binding_1D}).  Both $v_\eff$ and $k_\eff$ are important
to the low-energy effective theory of the $p$-wave resonance discussed
below.

\subsection{Low-energy effective theory}
The low-energy effective theory of the $p$-wave resonance in the 1D-3D
mixed dimensions is provided by the action
\begin{equation}\label{eq:action_1D}
 \begin{split}
  S &= \int\!dtdz\,\Psi_A^\+(t,z)
  \left(i\d_t+\frac{\nabla_{\!z}^2}{2m_A}\right)\Psi_A(t,z) \\
  &\quad + \int\!dtdzd\bm\rho\,\Psi_B^\+(t,z,\bm\rho)
  \left(i\d_t+\frac{\nabla_{\!z}^2+\grad_{\!\bm\rho}^2}{2m_B}\right)
  \Psi_B(t,z,\bm\rho) \\
  &\quad + \int\!dtdz\,\Phi_j^\+(t,z)
  \left(i\d_t+\frac{\nabla_{\!z}^2}{2M}+\varepsilon_0\right)\Phi_j(t,z) \\
  &\quad +g_0\int\!dtdz
  \left[\Psi_A^\+(t,z)\nabla_{\!j}\Psi_B^\+(t,z,\0)\Phi_j(t,z)\right. \\
  &\qquad\qquad\qquad
  \left.+\Phi_j^\+(t,z)\nabla_{\!j}\Psi_B(t,z,\0)\Psi_A(t,z)\right],
 \end{split}
\end{equation}
where the summation over $j=x,y$ is implicitly understood.  $\Psi_A$
and $\Psi_B$ fields represent the $A$ and $B$ atoms in 1D and 3D,
respectively.  The interaction between $A$ and $B$ atoms is described
through their coupling with two $p$-wave molecular fields $\Phi_j$.
$g_0$ is their coupling strength and $\varepsilon_0$ is the detuning
from the resonance.  These cutoff ($\Lambda$)-dependent bare parameters
can be related to the effective scattering volume $v_\eff$ and effective
momentum $k_\eff$ by matching the two-body scattering amplitude from the
action (\ref{eq:action_1D}) with that shown in
Eqs.~(\ref{eq:decomposition_1D}) and (\ref{eq:expansion_1D}).

The standard diagrammatic calculation leads to the following scattering
amplitude with collision energy $\varepsilon=k^2/(2m_B)$:
\begin{equation}
 \begin{split}
  i\mathcal{A}(k) = -\frac{i\k_\perp\cdot\k'_\perp}{\frac{k^2/(2m_B)+\varepsilon_0}{g_0^2}
  +\frac{\sqrt{m_Bm_{AB}}}{3\pi^2}\left(\frac{\Lambda^3}3+\Lambda k^2+\frac\pi2ik^3\right)}.
 \end{split}
\end{equation}
By defining
\begin{equation}
 \frac{\varepsilon_0}{g_0^2}+\frac{\sqrt{m_Bm_{AB}}}{3\pi^2}\frac{\Lambda^3}3
  \equiv \frac{\sqrt{m_Bm_{AB}}}{6\pi}\frac1{v_\eff}
\end{equation}
and
\begin{equation}
 \frac1{2m_Bg_0^2}+\frac{\sqrt{m_Bm_{AB}}}{3\pi^2}\Lambda
  \equiv -\frac{\sqrt{m_Bm_{AB}}}{6\pi}\frac{k_\eff}2,
\end{equation}
we reproduce the scattering amplitude (\ref{eq:expansion_1D}) in the
$p$-wave ($|m|=1$) channel up to a kinematical factor:
\begin{equation}
 \mathcal{A}(k) = -\frac{2\pi}{\sqrt{m_Bm_{AB}}}
  \frac{3\k_\perp\cdot\k'_\perp}{\frac1{v_\eff}-\frac{k_\eff}2k^2+ik^3}.
\end{equation}
This low-energy effective theory can be generalized easily to the case
with more than one tube where $B$ atoms are confined and could be used
to investigate the many-body physics across the $p$-wave resonance as in
the 3D case~\cite{Gurarie:2007}.  We note that the low-energy effective
theory of the $s$-wave resonance in the 1D-3D mixed dimensions has been
derived and used to study three-body problems in
Ref.~\cite{Nishida:2009fs}.

\subsection{Weak-coupling limit}
When $a_\mathrm{3D}<0$ and $|a_\mathrm{3D}|\ll l_\mathrm{ho}$, the
confinement-induced resonances can be understood in a different way.  To
the leading order in the weak-coupling expansion
$a_\mathrm{3D}/l_\mathrm{ho}\to-0$, an $A$ atom occupies the ground
state in a 2D harmonic potential and creates a mean-field attractive
potential felt by a $B$ atom.  Therefore, the scattering of the $B$ atom
by the confined $A$ atom is described by
\begin{equation}\label{eq:weak-coupling_1D}
 \begin{split}
  & \left[-\frac{\grad_{\!\bm\rho_B}^2}{2m_B}-\frac{\nabla_{\!z_{AB}}^2}{2m_{AB}}
  +\frac{2\pi a_\mathrm{3D}}{m_{AB}}|\phi_\0(\rho_B)|^2\delta(z_{AB})\right] \\
  &\times \psi(\bm\rho_B,z_{AB}) = \frac{k^2}{2m_B}\psi(\bm\rho_B,z_{AB}),
 \end{split}
\end{equation}
where
$|\phi_\0(\rho)|^2=e^{-\rho^2/l_\mathrm{ho}^2}/(\sqrt\pi\,l_\mathrm{ho})^2$.
This Schr\"odinger equation, which is valid in the weak-coupling limit,
is equivalent to the integral equation~(\ref{eq:chi_1D}), where only the
$\n=\0$ term is kept in the Green's function (\ref{eq:G_1D}) and
$\chi(\bm\rho)$ is identified as
$a_\mathrm{3D}\phi_\0(\rho)\psi(\bm\rho,0)$.

By matching the solution of Eq.~(\ref{eq:weak-coupling_1D}) with the
asymptotic form (\ref{eq:asymptotic_1D}), we can determine the
scattering amplitude and low-energy scattering parameters in the
weak-coupling limit $|a_\mathrm{3D}|\ll l_\mathrm{ho}$.  In particular,
the resonance occurs when a new bound state appears.  This is possible
even in the weak-coupling limit because the attractive potential becomes
strong compared to the kinetic term by decreasing the mass ratio down to
$m_A/m_B\ll1$.  We find that the resonances are achieved at the critical
values of
$\sqrt{m_B/m_{AB}}(a_\mathrm{3D}/l_\mathrm{ho})=-0.730,-2.55,-4.34,\dots$
in the $m=0$ channel,
$\sqrt{m_B/m_{AB}}(a_\mathrm{3D}/l_\mathrm{ho})=-1.96,-3.69,-5.44,\dots$
in the $m=1$ channel,
$\sqrt{m_B/m_{AB}}(a_\mathrm{3D}/l_\mathrm{ho})=-3.14,-4.85,-6.58,\dots$
in the $m=2$ channel, and
$\sqrt{m_B/m_{AB}}(a_\mathrm{3D}/l_\mathrm{ho})=-4.31,-6.00,-7.72,\dots$
in the $m=3$ channel.

\section{2D-3D mixed dimensions \label{sec:2D-3D}}

\subsection{Scattering theory~\cite{normalization}}
The scattering of a quasi-2D $A$ atom with a $B$ atom in 3D is described
by a Schr\"odinger equation,
\begin{equation}\label{eq:schrodinger_2D}
 \begin{split}
  & \left(-\frac{\nabla_{\!z_A}^2}{2m_A}+\frac12m_A\omega^2z_A^2
  -\frac{\nabla_{\!z_B}^2}{2m_B}-\frac{\grad_{\!\bm\rho_{AB}}^2}{2m_{AB}}\right) \\
  &\times \psi(z_A,z_B,\bm\rho_{AB}) = E\,\psi(z_A,z_B,\bm\rho_{AB})
 \end{split}
\end{equation}
for $\sqrt{(z_A-z_B)^2+\bm\rho_{AB}^2}>0$, where
$\bm\rho_{AB}\equiv(x_A-x_B,y_A-y_B)$ and the center-of-mass motions in
the $x$ and $y$ directions are eliminated.  The short-range interspecies
interaction is implemented by the generalized Bethe-Peierls boundary
condition~\cite{Petrov:2004,Levinsen:2009mn}:
\begin{equation}\label{eq:short-range_2D}
 \begin{split}
  & \psi(z_A,z_B,\bm\rho_{AB})\big|_{z_A,z_B\to z;\bm\rho_{AB}\to\0} \\
  &\to \left[\frac1{\tilde a(\hat E_c)}
  -\frac1{\sqrt{(z_A-z_B)^2+\bm\rho_{AB}^2}}\right]\chi(z).
 \end{split}
\end{equation}
The collision energy operator $\hat E_c$ in Eq.~(\ref{eq:generalized_a})
in the present case is given by
\begin{equation}
 \hat E_c = E
  - \left(-\frac{\nabla_{\!z}^2}{2M}+\frac12m_A\omega^2z^2\right).
\end{equation}

The solution to the Schr\"odinger equation (\ref{eq:schrodinger_2D}) can
be written as
\begin{equation}\label{eq:solution_2D}
 \begin{split}
  & \psi(z_A,z_B,\bm\rho_{AB}) = \psi_0(z_A,z_B,\bm\rho_{AB}) \\
  &\quad + \frac{2\pi}{m_{AB}}\int\!dz'
  G_E(z_A,z_B,\bm\rho_{AB};z',z',\0)\chi(z'),
 \end{split}
\end{equation}
where $\psi_0$ is a solution in the noninteracting limit and $G_E$ is the
retarded Green's function for the noninteracting Hamiltonian:
\begin{equation}\label{eq:G_2D}
 \begin{split}
  & G_E(z_A,z_B,\bm\rho_{AB};z'_A,z'_B,\bm\rho'_{AB}) \\
  &\equiv \<z_A,z_B,\bm\rho_{AB}|\frac1{E-H_0+i0^+}|z'_A,z'_B,\bm\rho'_{AB}\> \\
  &= -\frac{m_{AB}}{2\pi}\sum_{n_z}\phi_{n_z}(z_A)\phi_{n_z}^*(z'_A) \\
  &\quad \times \frac{e^{-\sqrt{2m_B}\sqrt{(n_z+\frac12)\omega-E-i0^+}
  |\tilde\r_B-\tilde\r_B'|}}{|\tilde\r_B-\tilde\r_B'|}.
 \end{split}
\end{equation}
Here $\phi_{n_z}$ is the normalized wave function of an $A$ atom in the
1D harmonic potential, and
$\tilde\r_B\equiv\left(-\sqrt\frac{m_{AB}}{m_B}\,\bm\rho_{AB},z_B\right)$
are coordinates of the $B$ atom relative to the confined $A$ atom.  The
anisotropic factor is such because a separation in the $z$ direction is
associated with $m_B$ while separations in the $x$ and $y$ directions
are associated with $m_{AB}$ [see the last two terms in
Eq.~(\ref{eq:schrodinger_2D})].

We now consider the low-energy scattering in which
\begin{equation}
 E-\frac12\omega \equiv \frac{k^2}{2m_B} \ll \omega
\end{equation}
is satisfied, and then $\psi_0$ becomes
\begin{equation}
 \psi_0(z_A,z_B,\bm\rho_{AB}) = Ce^{i\k\cdot\tilde\r_B}\phi_0(z_A),
\end{equation}
which represents the $A$ atom in the ground state of the 1D harmonic
potential and the plane wave of $B$ atom with the wave vector $\k$.  The
asymptotic form of the wave function at a large distance
$|\tilde\r_B|\gg l_\mathrm{ho}$ is given by
\begin{equation}\label{eq:asymptotic_2D}
 \psi(z_A,z_B,\bm\rho_{AB}) \to C\left[e^{i\k\cdot\tilde\r_B}
  +\frac{e^{ik\tilde r_B}}{\tilde r_B}f(k_z,k'_z)\right]\phi_0(z_A),
\end{equation}
where $f(k_z,k'_z)$ with $\k'\equiv k\hat{\tilde\r}_B$ defines the
two-body scattering amplitude in the 2D-3D mixed dimensions:
\begin{equation}\label{eq:f_2D}
 f(k_z,k'_z) \equiv -\frac1{C}\int\!dz'e^{-ik'_zz'}\phi_0^*(z')\chi(z').
\end{equation}
We note that $\chi$ has an implicit $k_z$ dependence and both $\chi$ and
$f$ depend on $k$ through the Green's function (\ref{eq:G_2D}).

The unknown function $\chi$ can be determined by substituting the
solution (\ref{eq:solution_2D}) into the Bethe-Peierls boundary
condition (\ref{eq:short-range_2D}).  Defining the regular part of the
Green's function $\G$ by
\begin{equation}\label{eq:regular_2D}
 \begin{split}
  & G_E(z_A,z_B,\bm\rho_{AB};z',z',\0)\big|_{z_A,z_B\to z;\bm\rho_{AB}\to\0} \\
  &\equiv -\frac{m_{AB}}{2\pi\sqrt{(z_A-z_B)^2+\bm\rho_{AB}^2}}\delta(z-z') + \G(z;z'),
 \end{split}
\end{equation}
we obtain
\begin{equation}\label{eq:chi_2D}
 \frac1{\tilde a(\hat E_c)}\chi(z) = C e^{ik_zz}\phi_0(z)
  + \frac{2\pi}{m_{AB}}\int\!dz'\G(z;z')\chi(z').
\end{equation}
This integral equation determines $\chi/C$, which in turn provides $f$
from Eq.~(\ref{eq:f_2D}).

\begin{figure*}[t]\hfill
 \includegraphics[width=0.46\textwidth,clip]{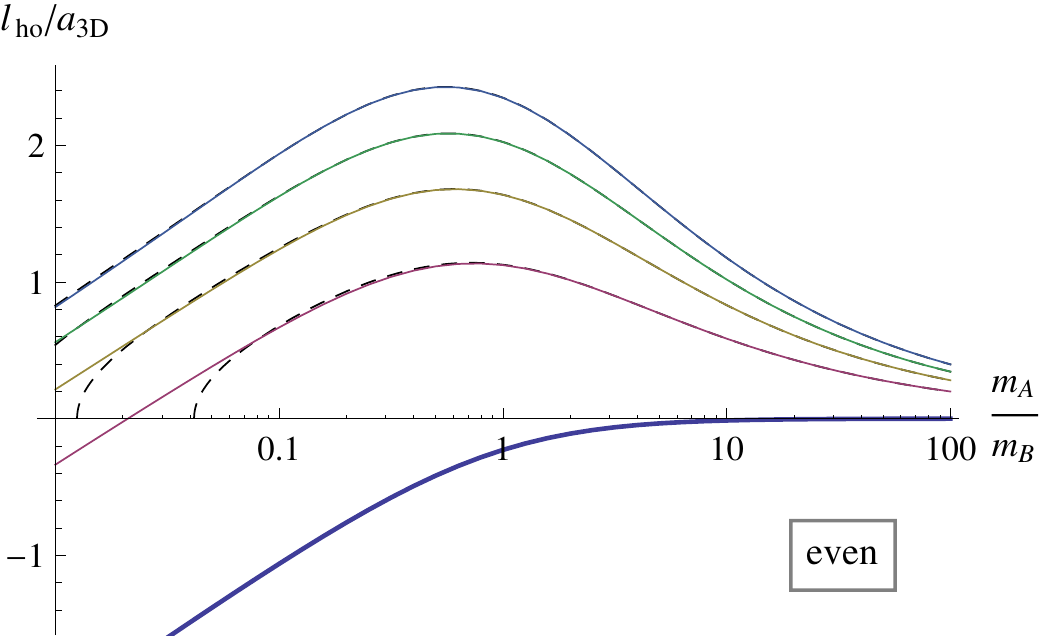}\hfill\hfill\hfill
 \includegraphics[width=0.46\textwidth,clip]{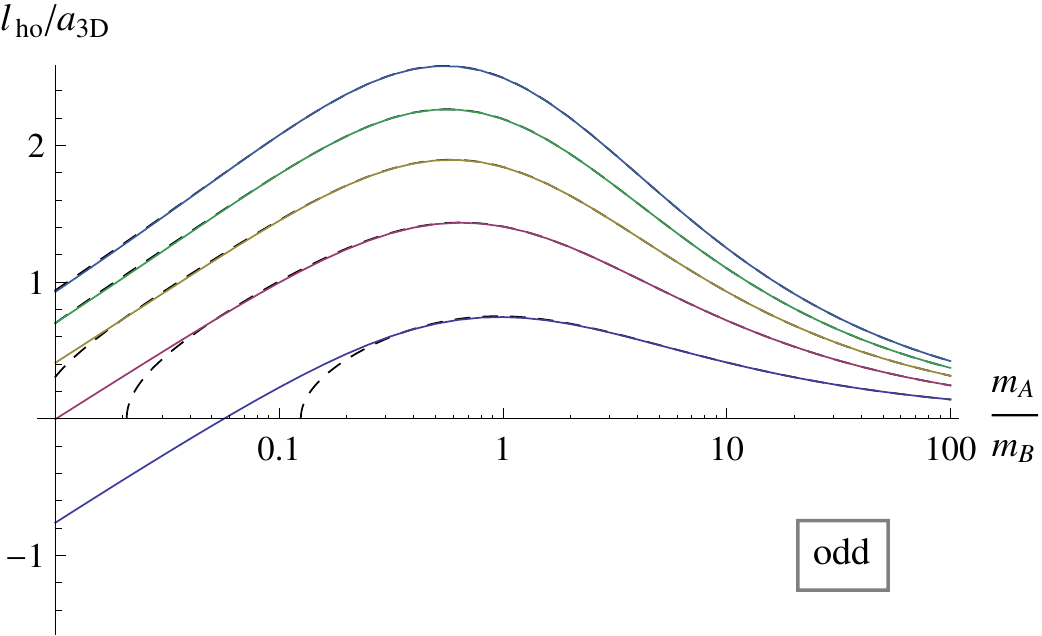}\hfill\hfill
 \caption{(Color online) 2D-3D mixture: Positions of the lowest five
 resonances in terms of $l_\mathrm{ho}/a_\mathrm{3D}$ for even-parity
 (left) and odd-parity (right) channels as functions of the mass ratio
 $m_A/m_B$.  The dashed curves are from the approximate formula
 $E_{AB}=\frac12\hbar\omega$ by using Eq.~(\ref{eq:approx_2D}) with
 $E_b=-\hbar^2/(2m_{AB}a_\mathrm{3D}^2)$.  \label{fig:resonance_2D}}
\end{figure*}

Because the system has a reflection symmetry about the $z$ axis, $f$,
$\chi$, and $\G$ can be decomposed into their even- and odd-parity
components:
\begin{align}\label{eq:decomposition_2D}
 f_\pm(k_z,k'_z) &= \frac{f(k_z,k'_z)\pm f(k_z,-k'_z)}2, \\
 \chi_\pm(z) &= \frac{\chi(z)\pm\chi(-z)}2, \\
 \G_\pm(z;z') &= \frac{\G(z;z')\pm\G(z;-z')}2.
\end{align}
Equations~(\ref{eq:f_2D}) and (\ref{eq:chi_2D}) lead to the
even- and odd-parity scattering amplitudes given by
\begin{equation}\label{eq:partial-wave_2D}
 f_\pm(k_z,k'_z) = -\frac1{C}\int\!dz'\frac{e^{-ik'_zz'}\pm e^{ik'_zz'}}2
  \phi_0^*(z')\chi_\pm(z')
\end{equation}
with
\begin{equation}
 \begin{split}
  \frac1{\tilde a(\hat E_c)}\chi_\pm(z) &= C\frac{e^{ik_zz}\pm e^{-ik_zz}}2\phi_0(z) \\
  &\quad + \frac{2\pi}{m_{AB}}\int\!dz'\G_\pm(z;z')\chi_\pm(z').
 \end{split}
\end{equation}
From the explicit calculation that uses the Green's function in
Eq.~(\ref{eq:G_2D}), we can show that $f_\pm$ has the following
low-energy expansion:
\begin{equation}\label{eq:expansion_2D_even}
 \begin{split}
  & \lim_{k\to0}f_+(k_z,k'_z) \\
  &= -\frac1{\frac1{a_\eff^{(+)}}
  -\frac12r_\eff^{(+)}k^2+O(k^4)+ik\left[1+O(k^2)\right]} \\
  &\quad + O(k^2_z,k'^2_z)
 \end{split}
\end{equation}
and
\begin{equation}\label{eq:expansion_2D_odd}
 \begin{split}
  & \lim_{k\to0}f_-(k_z,k'_z) \\
  &= -\frac{3k_zk'_z}{\frac1{a_\eff^{(-)}}
  -\frac12r_\eff^{(-)}k^2+O(k^4)+ik^3\left[1+O(k^2)\right]} \\
  &\quad + O(k^3_zk'_z,k_zk'^3_z),
 \end{split}
\end{equation}
where $a_\eff^{(\pm)}$ and $r_\eff^{(\pm)}$ are effective scattering
``length'' and ``range'' parameters in the even- and odd-parity
channels.  Note that $a_\eff^{(\pm)}$ has the dimension of
$(\mathrm{length})^{2\mp1}$ and $r_\eff^{(\pm)}$ has
$(\mathrm{length})^{\pm1}$.  This unusual form of the low-energy
expansion is owing to the lack of full 3D rotational symmetries.

Substituting the expansion of $f_\pm$ into
Eq.~(\ref{eq:partial-wave_2D}), we can determine the low-energy
scattering parameters.  In particular, the effective scattering length
$a_\eff^{(\pm)}$ is given by
\begin{equation}\label{eq:a_eff_2D}
 \begin{split}
  \frac1{\tilde a(\hat E_c)}\chi_+(z) 
  &= \frac1{a_\eff^{(+)}}\phi_0(z)\int\!dz'\phi_0^*(z')\chi_+(z') \\
  &\quad + \frac{2\pi}{m_{AB}}\int\!dz'\G_+(z;z')\big|_{k\to0}\chi_+(z')
 \end{split}
\end{equation}
and
\begin{equation}\label{eq:v_eff_2D}
 \begin{split}
  \frac1{\tilde a(\hat E_c)}\chi_-(z) 
  &= \frac1{3a_\eff^{(-)}}z\phi_0(z)\int\!dz'z'\phi_0^*(z')\chi_-(z') \\
  &\quad + \frac{2\pi}{m_{AB}}\int\!dz'\G_-(z;z')\big|_{k\to0}\chi_-(z'),
 \end{split}
\end{equation}
where we have eliminated $C$.  The even- and odd-parity resonances in
the 2D-3D mixed dimensions are defined by the divergence of
$a_\eff^{(\pm)}\to\infty$, which occurs when
\begin{equation}\label{eq:resonance_2D}
 \frac1{\tilde a(\hat E_c)}\chi_\pm(z) 
  = \frac{2\pi}{m_{AB}}\int\!dz'\G_\pm(z;z')\big|_{k\to0}\chi_\pm(z')
\end{equation}
is satisfied.

\subsection{Positions of resonances}
We now solve the integral equation (\ref{eq:resonance_2D}) numerically
to determine the positions of even- and odd-parity resonances.  For the
purpose of illustrating qualitative results, we shall set
$r_\mathrm{3D}=0$.  For quantitative predictions in a specific atomic
mixture, it is necessary but straightforward to include the effective
range correction~\cite{Lamporesi:2010}.  Some details of our method to
solve the integral equation are shown in Appendix~\ref{app:2D-3D}.

Figure~\ref{fig:resonance_2D} shows the positions of the lowest five
resonances in terms of $l_\mathrm{ho}/a_\mathrm{3D}$ for even- and
odd-parity channels as functions of the mass ratio $m_A/m_B$.  For
completeness, we have included the result for the even-parity resonance,
which has been reported in Ref.~\cite{Nishida:2008kr}.  As we have
discussed in Sec.~\ref{sec:introduction}, there exists a series of
resonances in each parity channel induced from the purely $s$-wave
interaction in a free space.  Indeed, the resonance positions are well
described by the approximate formula $E_{AB}=\frac12\hbar\omega$ by
using Eq.~(\ref{eq:approx_2D}) with
$E_b=-\hbar^2/(2m_{AB}a_\mathrm{3D}^2)$ in a wide range of the mass
ratio $m_A/m_B\gtrsim1$.

\begin{figure*}[t]
 \begin{minipage}[t]{0.33\textwidth}
  \begin{center}\fbox{\scriptsize$m_A/m_B=6/40$}\end{center}
  \includegraphics[width=\textwidth,clip]{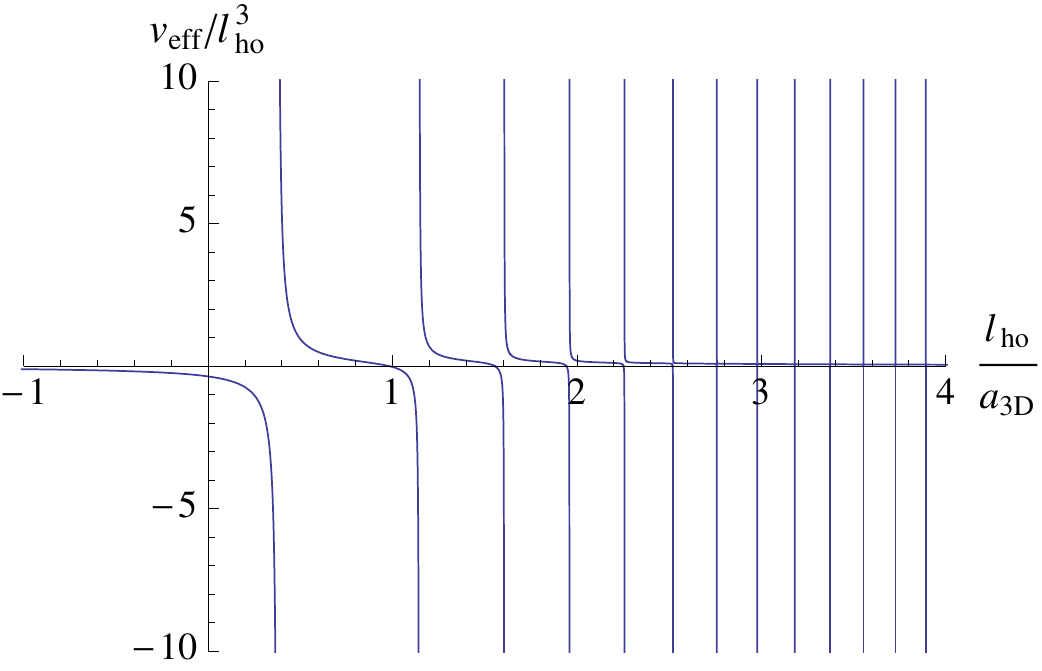}\vspace{4mm}
  \includegraphics[width=\textwidth,clip]{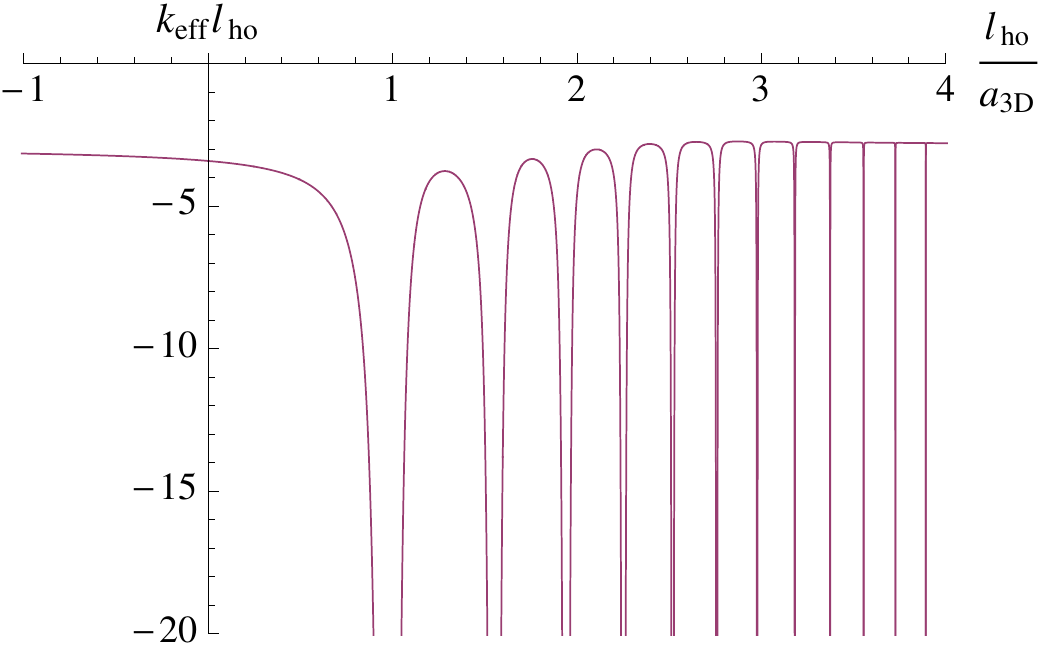}
 \end{minipage}\hfill
 \begin{minipage}[t]{0.33\textwidth}
  \begin{center}\fbox{\scriptsize$m_A/m_B=1$}\end{center}
  \includegraphics[width=\textwidth,clip]{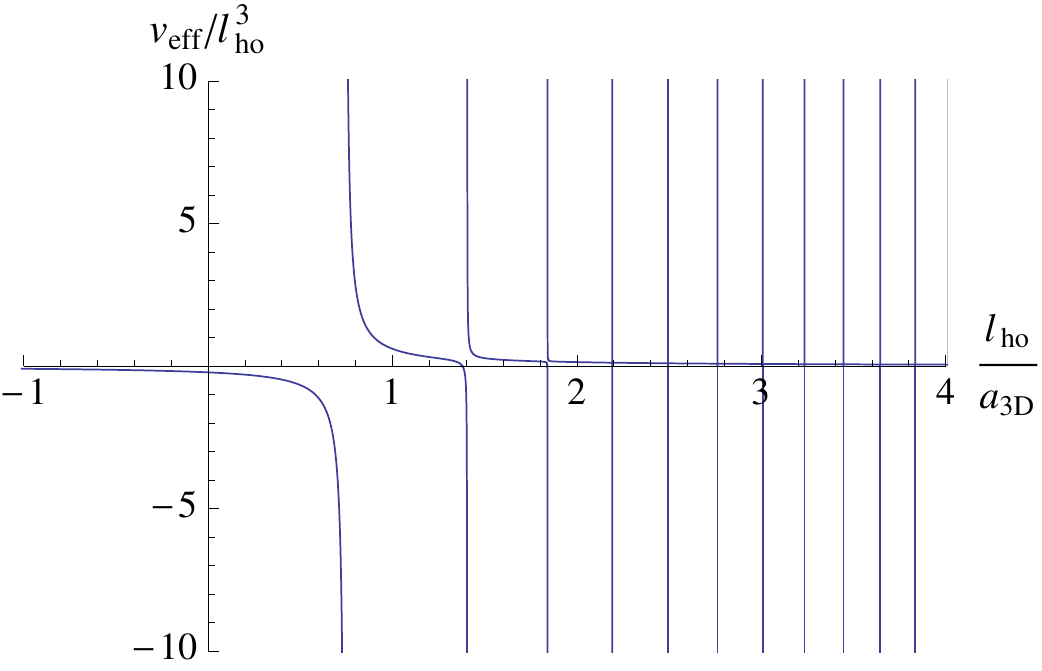}\vspace{4mm}
  \includegraphics[width=\textwidth,clip]{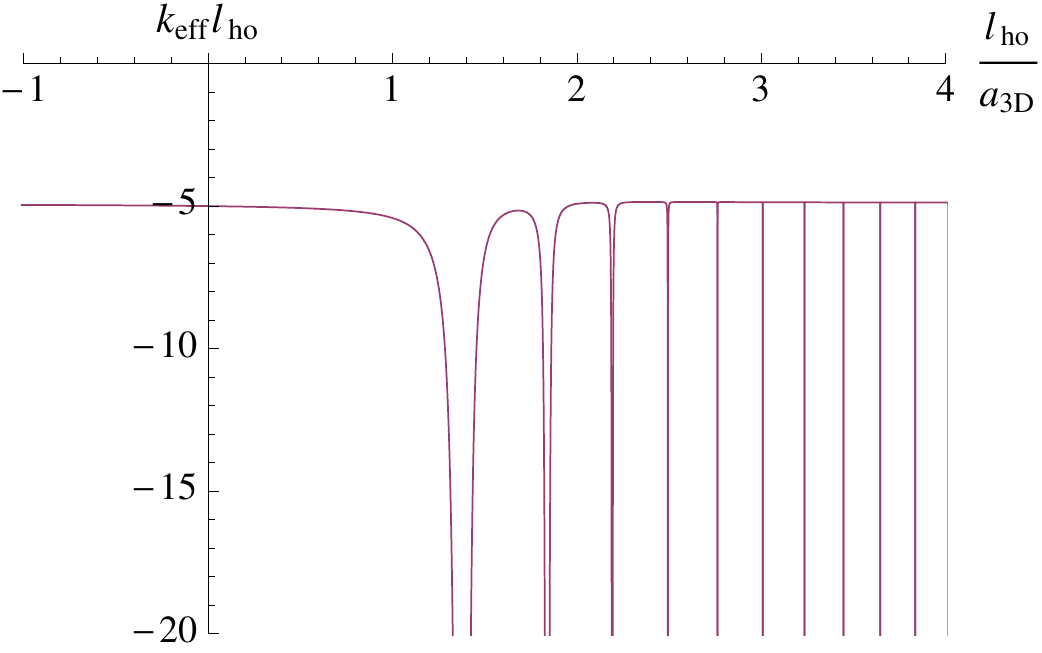}
 \end{minipage}\hfill
 \begin{minipage}[t]{0.33\textwidth}
  \begin{center}\fbox{\scriptsize$m_A/m_B=40/6$}\end{center}
  \includegraphics[width=\textwidth,clip]{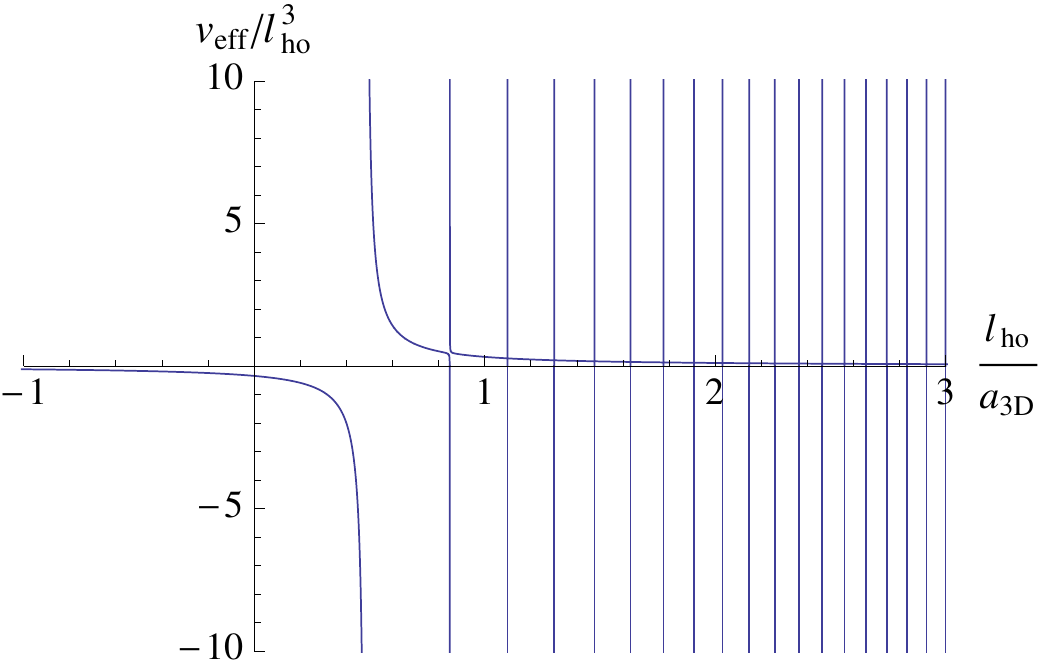}\vspace{4mm}
  \includegraphics[width=\textwidth,clip]{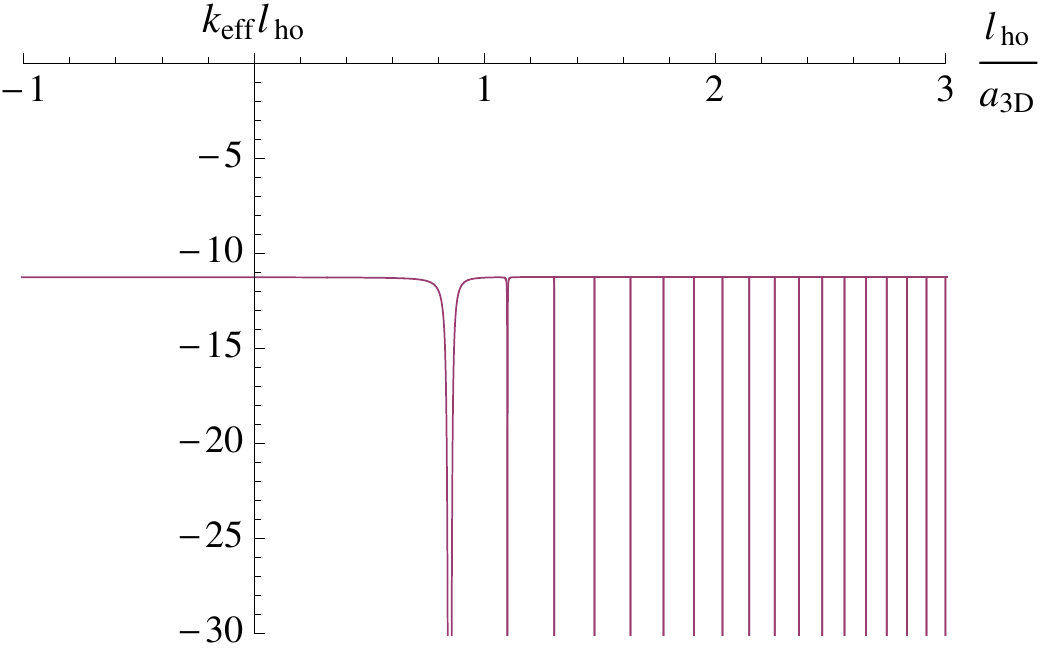}
 \end{minipage}
 \caption{(Color online) 2D-3D mixture: $p$-wave effective scattering
 volume $v_\eff/l_\mathrm{ho}^3$ (upper figures) and effective momentum
 $k_\eff l_\mathrm{ho}$ (lower figures) as functions of
 $l_\mathrm{ho}/a_\mathrm{3D}$ for mass ratios $m_A/m_B=6/40$ (left),
 $1$ (middle), and $40/6$ (right).  \label{fig:parameter_2D}}
\end{figure*}

\subsection{Confinement-induced molecules}
On the $a_\eff^{(\pm)}>0$ side of every resonance, a shallow $AB$
molecule is formed.  In the vicinity of the resonance
$a_\eff^{(\pm)}\gg l_\mathrm{ho}^{2\mp1}$, its binding energy
$\varepsilon_{AB}\equiv E-\frac12\omega<0$ is determined by the pole of
the scattering amplitude [Eqs.~(\ref{eq:expansion_2D_even}) and
(\ref{eq:expansion_2D_odd})] with keeping the two dominant terms at
$k\to0$:
\begin{equation}\label{eq:binding_2D}
 \varepsilon_{AB} =
  \begin{cases}
   \displaystyle -\frac1{2m_Ba_\eff^{(+)2}} 
   &\ \text{for even parity}, \medskip\\
   \displaystyle \phantom{\,}\frac1{m_Ba_\eff^{(-)}r_\eff^{(-)}}
   &\ \text{for odd parity}.
  \end{cases}
\end{equation}
Away from the resonance, these universal formulas are no longer valid.
The binding energy $\varepsilon_{AB}=-\kappa^2/(2m_B)$ has to be
determined by solving the integral equation
\begin{equation}
 \frac1{\tilde a(\hat E_c)}\chi_\pm(z) = \frac{2\pi}{m_{AB}}
  \int\!dz'\G_\pm(z;z')\big|_{k\to i\kappa}\chi_\pm(z').
\end{equation}

We now derive the asymptotic form of the molecular wave function in the
vicinity of the even- or odd-parity resonance where
$l_\mathrm{ho}\ll\kappa^{-1}\ll|\tilde\r_B|$ is satisfied.  From
Eqs.~(\ref{eq:solution_2D}) and (\ref{eq:G_2D}) with the replacement
$k\to i\kappa$, we find
\begin{equation}
 \begin{split}
  \psi(z_A,z_B,\bm\rho_{AB})
  &\to -\sum_{\ell=\text{even or odd}}^\infty\sqrt\frac{4\pi}{2\ell+1}
  \frac{e^{-\kappa\tilde r_B}}{\tilde r_B^{\ell+1}}\phi_0(z_A) \\
  &\quad \times Y_\ell^0(\hat{\tilde\r}_B)
  \int\!dz'|z'|^\ell\phi_0^*(z')\chi_\pm(z').
 \end{split}
\end{equation}
One can see that different spherical harmonics $\sim Y_\ell^0$
contribute owing to the lack of full 3D rotational symmetries.  The
$s$-wave nature of the free-space interaction ensures that only the
$m=0$ component contributes so that the wave function is independent of
$\hat{\bm\rho}_{AB}$.  The asymptotic behavior at a large separation
$|\tilde\r_B|\to\infty$ is dominated by the $\ell=0$ or $1$ component in
the even- or odd-parity channel, respectively.  Therefore, we can phrase
the even (odd)-parity resonance as an $s$-wave ($p$-wave) resonance in
the 2D-3D mixture.  The angular part of the asymptotic wave function of
a shallow $p$-wave molecule $\psi\sim Y_1^0(\hat{\tilde\r}_B)$ is
illustrated in Fig.~\ref{fig:molecule}.

\subsection{Scattering parameters in the $p$-wave channel}
The effective scattering length in the $s$-wave (even-parity) channel
$a_\eff\equiv a_\eff^{(+)}$ has been computed in
Ref.~\cite{Nishida:2008kr}.  Here we focus on the $p$-wave (odd-parity)
channel and determine its two low-energy scattering parameters, namely,
the effective scattering volume $v_\eff\equiv a_\eff^{(-)}$ and the
effective momentum $k_\eff\equiv r_\eff^{(-)}$.  The effective
scattering volume can be computed by solving the integral equation
(\ref{eq:v_eff_2D}) numerically (see Appendix~\ref{app:2D-3D} for
details).  In Fig.~\ref{fig:parameter_2D}, $v_\eff/l_\mathrm{ho}^3$ for
$r_\mathrm{3D}=0$ is plotted as a function of
$l_\mathrm{ho}/a_\mathrm{3D}$ for three mass ratios $m_A/m_B=6/40$, $1$,
and $40/6$.  We confirm the existence of a series of $p$-wave resonances
($v_\eff\to\infty$) induced from the purely $s$-wave interaction in a
free space, while they become narrower for larger
$l_\mathrm{ho}/a_\mathrm{3D}$.  We also find that the resonance is wider
when a lighter atom is confined in lower dimensions.

Similarly, the effective momentum can be computed from
Eq.~(\ref{eq:partial-wave_2D}), and $k_\eff l_\mathrm{ho}$ for
$r_\mathrm{3D}=0$ is plotted in Fig.~\ref{fig:parameter_2D} as a
function of $l_\mathrm{ho}/a_\mathrm{3D}$ for the same three mass
ratios.  In the vicinity of the $p$-wave resonance
$v_\eff\gg l_\mathrm{ho}^3$, $v_\eff$ and $k_\eff$ determine the binding
energy of a shallow $p$-wave molecule via the universal formula
(\ref{eq:binding_2D}).  Both $v_\eff$ and $k_\eff$ are important to the
low-energy effective theory of the $p$-wave resonance discussed below.

\subsection{Low-energy effective theory}
The low-energy effective theory of the $p$-wave resonance in the 2D-3D
mixed dimensions is provided by the action
\begin{equation}\label{eq:action_2D}
 \begin{split}
  S &= \int\!dtd\bm\rho\,\Psi_A^\+(t,\bm\rho)
  \left(i\d_t+\frac{\grad_{\!\bm\rho}^2}{2m_A}\right)\Psi_A(t,\bm\rho) \\
  &\quad + \int\!dtd\bm\rho dz\,\Psi_B^\+(t,\bm\rho,z)
  \left(i\d_t+\frac{\grad_{\!\bm\rho}^2+\nabla_{\!z}^2}{2m_B}\right)
  \Psi_B(t,\bm\rho,z) \\
  &\quad + \int\!dtd\bm\rho\,\Phi^\+(t,\bm\rho)
  \left(i\d_t+\frac{\grad_{\!\bm\rho}^2}{2M}+\varepsilon_0\right)\Phi(t,\bm\rho) \\
  &\quad + g_0\int\!dtd\bm\rho\left[\Psi_A^\+(t,\bm\rho)
  \nabla_{\!z}\Psi_B^\+(t,\bm\rho,0)\Phi(t,\bm\rho)\right. \\
  &\qquad\qquad\qquad \left.+\Phi^\+(t,\bm\rho)
  \nabla_{\!z}\Psi_B(t,\bm\rho,0)\Psi_A(t,\bm\rho)\right].
 \end{split}
\end{equation}
$\Psi_A$ and $\Psi_B$ fields represent the $A$ and $B$ atoms in 2D and
3D, respectively.  The interaction between $A$ and $B$ atoms is described
through their coupling with a $p$-wave molecular field $\Phi$.  $g_0$ is
their coupling strength and $\varepsilon_0$ is the detuning from the
resonance.  These cutoff ($\Lambda$)-dependent bare parameters can be
related to the effective scattering volume $v_\eff$ and effective
momentum $k_\eff$ by matching the two-body scattering amplitude from the
action (\ref{eq:action_2D}) with that shown in
Eqs.~(\ref{eq:decomposition_2D}) and (\ref{eq:expansion_2D_odd}).

The standard diagrammatic calculation leads to the following scattering
amplitude with collision energy $\varepsilon=k^2/(2m_B)$:
\begin{equation}
 \begin{split}
  i\mathcal{A}(k) = -\frac{ik_zk'_z}{\frac{k^2/(2m_B)+\varepsilon_0}{g_0^2}
  +\frac{m_{AB}}{3\pi^2}\left(\frac{\Lambda^3}3+\Lambda k^2+\frac\pi2ik^3\right)}.
 \end{split}
\end{equation}
By defining
\begin{equation}
 \frac{\varepsilon_0}{g_0^2}+\frac{m_{AB}}{3\pi^2}\frac{\Lambda^3}3
  \equiv \frac{m_{AB}}{6\pi}\frac1{v_\eff}
\end{equation}
and
\begin{equation}
 \frac1{2m_Bg_0^2}+\frac{m_{AB}}{3\pi^2}\Lambda
  \equiv -\frac{m_{AB}}{6\pi}\frac{k_\eff}2,
\end{equation}
we reproduce the scattering amplitude (\ref{eq:expansion_2D_odd}) in the
$p$-wave (odd-parity) channel up to a kinematical factor:
\begin{equation}
 \mathcal{A}(k) = -\frac{2\pi}{m_{AB}}
  \frac{3k_zk'_z}{\frac1{v_\eff}-\frac{k_\eff}2k^2+ik^3}.
\end{equation}
This low-energy effective theory can be generalized easily to the case
with more than one layer where $B$ atoms are confined and could be used
to investigate the many-body physics across the $p$-wave resonance as in
the 3D case~\cite{Gurarie:2007}.  We note that the low-energy effective
theory of the $s$-wave resonance in the 2D-3D mixed dimensions has been
derived and used to study many-body problems in
Refs.~\cite{Nishida:2008gk,Nishida:2009nc}.

\subsection{Comparison to experiment~\cite{Lamporesi:2010}}
\begin{figure}[t]
 \includegraphics[width=0.98\columnwidth,clip]{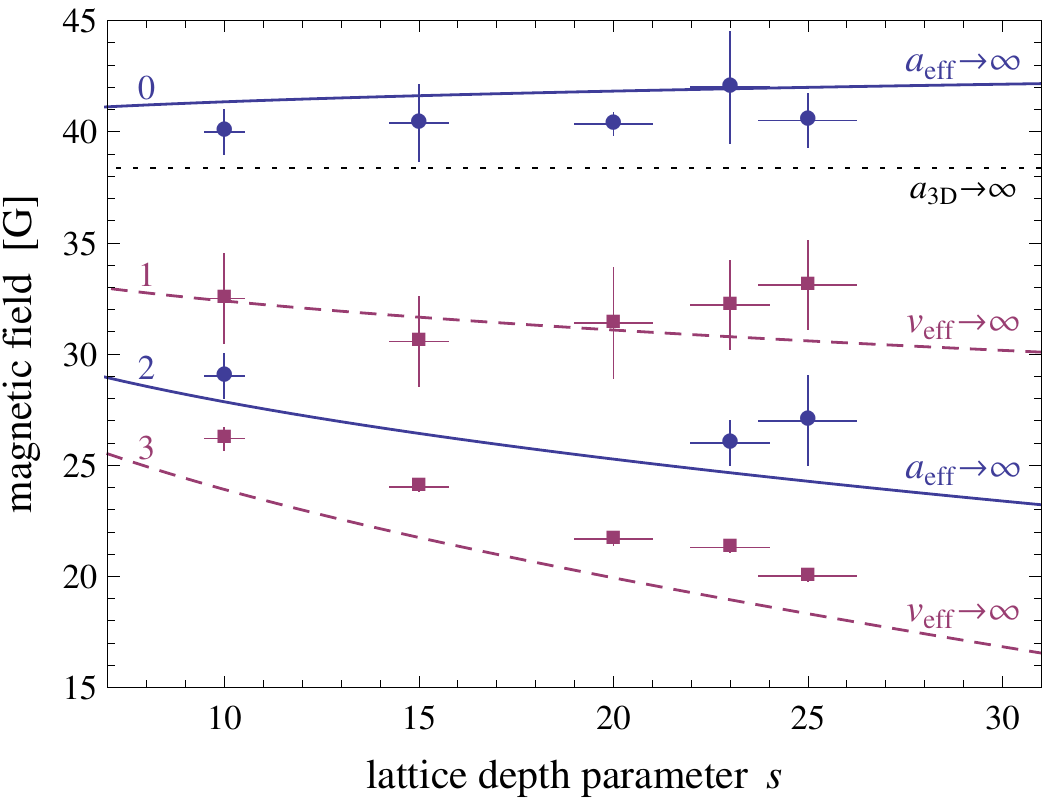}
 \caption{(Color online) Experimentally measured resonance positions in
 terms of the magnetic field $B$ as functions of the depth parameter $s$
 of an optical lattice~\cite{Lamporesi:2010}.  Also shown are the
 predicted positions of $s$-wave ($a_\eff\to\infty$; solid curves) and
 $p$-wave ($v_\eff\to\infty$; dashed curves) resonances in a harmonic
 potential as well as the resonance position in a free space
 ($a_\mathrm{3D}\to\infty$; dotted line).  \label{fig:experiment}}
\end{figure}

Finally, we compare our predictions to the experimental measurement of
resonance positions reported in Ref.~\cite{Lamporesi:2010}.  The
Florence group has realized the 2D-3D mixed dimensions by using a
Bose-Bose mixture of $A=^{41}$K and $B=^{87}$Rb with a species-selective
1D optical lattice:
\begin{equation}\label{eq:lattice}
 V(z_A) = s\frac{k_L^2}{2m_A}\sin^2(k_Lz_A).
\end{equation}
Here $s$ is the lattice depth parameter and $k_L=2\pi/\lambda_L$ with
$\lambda_L=790.02$\,nm is the wave vector of the laser light.  By
monitoring three-body inelastic losses, a series of resonances has been
observed as a function of the magnetic field.  The measured resonance
positions are shown in Fig.~\ref{fig:experiment} for five values of the
lattice depth parameter, $s=10,\,15,\,20,\,23$, and $25$.

As for theoretical predictions, we assume that the lattice potential
(\ref{eq:lattice}) can be approximated by a harmonic potential:
\begin{equation}\label{eq:harmonic}
 V(z_A) \approx s\frac{k_L^4}{2m_A}z_A^2
  \equiv \frac12m_A\omega^2z_A^2.
\end{equation}
The harmonic oscillator length is given from the laser wavelength by
$l_\mathrm{ho}=\lambda_L/(2\pi s^{1/4})$.  With the use of the
free-space value of the effective range $r_\mathrm{3D}=168.37\,a_0$ for
the $^{41}$K-$^{87}$Rb mixture, we solve the integral equation
(\ref{eq:resonance_2D}) numerically to determine the resonance positions
in terms of $l_\mathrm{ho}/a_\mathrm{3D}$.  The free-space scattering
length $a_\mathrm{3D}$ thus obtained is converted into the magnetic
field $B$\,(G) by the following empirical formula~\cite{Lamporesi:2010}:
\begin{equation}
 \frac{a_\mathrm{3D}}{a_0} = 208
  \left(1+\frac{30.9}{B+38.52}-\frac{49.92}{B-38.37}-\frac{1.64}{B-78.67}\right),
\end{equation}
where $a_0$ is the Bohr radius and the resonance in a free space occurs
at $B=38.4$\,G.  These critical magnetic field values for the $s$-wave
($a_\eff\to\infty$) and $p$-wave ($v_\eff\to\infty$) resonances are
plotted in Fig.~\ref{fig:experiment} as functions of the lattice depth
parameter $s$.

In Fig.~\ref{fig:experiment}, the resonances are labeled by an integer
$n=0,1,2,3$ in descending order, corresponding to $n$ in
Eq.~(\ref{eq:approx_2D}).  One can see the reasonable agreement between
the experimental measurements and the theoretical predictions for $n=0$
($s$-wave) and $n=1$ ($p$-wave) resonances.  They begin to deviate for
larger $n$ simply because the harmonic potential approximation
(\ref{eq:harmonic}) becomes worse for higher excited states in an
optical lattice.  Although the lattice potential (\ref{eq:lattice}) has
to be taken into account to improve the quantitative
agreement~\cite{Lamporesi:2010}, for a sufficiently strong optical
lattice, the resonances corresponding to odd $n$ should be the $p$-wave
resonances in the 2D-3D mixed dimensions, as we have discussed in this
section.  Further experimental investigations to confirm the $p$-wave
nature of these resonances would be worthwhile.  The $p$-wave nature of
shallow molecules as illustrated in Fig.~\ref{fig:molecule} could be
seen as a suppression of the density of $B$ atoms, as opposed to an
enhancement for $s$-wave molecules, at the center of the confinement
potential after a sudden change of $v_\eff>0$ to its negative side.

\section{Summary and concluding remarks \label{sec:summary}}
In this paper, we showed that a purely $s$-wave interaction in a free
space can induce higher partial-wave resonances in mixed dimensions.  We
developed two-body scattering theories in all three cases of 0D-3D,
1D-3D, and 2D-3D mixtures, and determined the positions of higher
partial-wave resonances in terms of the free-space scattering length,
assuming a harmonic confinement potential.  We also computed the
low-energy scattering parameters in the $p$-wave channel (effective
scattering volume and momentum) that are necessary for the low-energy
effective theory of the $p$-wave resonance.  Potentially our study paves
the way for a variety of physics, such as Anderson localization of matter
waves under $p$-wave resonant scatterers in the 0D-3D mixed
dimensions~\cite{Gavish:2005,Massignan:2006,Antezza:2010}, Shiba bound
states in a superfluid medium~\cite{Vernier:2010}, and Bose-Einstein
condensation of $p$-wave molecules in the 2D-3D mixed dimensions,
provided that such molecules are long lived.

When both $A$ and $B$ atoms are fermionic (such as for the
$^6$Li-$^{40}$K
mixture~\cite{Taglieber:2008,Wille:2008,Voigt:2009,Tiecke:2010,Naik:2010}),
inelastic three-body, atom-molecule, and molecule-molecule collisions
decaying into deeply bound dimers whose size is set by the range of
interatomic potential $r_0$ are suppressed.  This is because at a short
distance $\sim r_0\ll l_\mathrm{ho}$, the confinement potential is
irrelevant, and in the $s$-wave interspecies interaction the Pauli
exclusion principle is effective to suppress the inelastic collisions
decaying into the deeply bound dimers~\cite{Petrov:2003}.  An
order-of-magnitude estimate of such a three-body recombination rate can
be found in Refs.~\cite{Nishida:2009fs} and \cite{Levinsen:2009mn} for
wide Feshbach resonances and in Ref.~\cite{Levinsen:2009mn} for narrow
Feshbach resonances.

Therefore, for a Fermi-Fermi mixture in mixed dimensions, the relaxation
of molecules is dominated by the decay into deeper molecular states of
size $\sim l_\mathrm{ho}$ if they exist.  In the vicinity of the
$s$-wave resonance at the smallest value of
$l_\mathrm{ho}/a_\mathrm{3D}$, indicated by thick curves in
Figs.~\ref{fig:resonance_0D}, \ref{fig:resonance_1D}, and
\ref{fig:resonance_2D}, there is no such molecular state and thus
associated $s$-wave molecules are long lived.  However, in the vicinity
of the higher partial-wave resonances, there is always at least one
deeper molecular state.  The inelastic atom-molecule and
molecule-molecule collisions decaying into such molecular states are not
generally suppressed, and thus $p$-wave or higher partial-wave molecules
in mixed dimensions would be short lived.  It is therefore an important
future problem to investigate as to whether the inelastic collisions of
$p$-wave molecules decaying into the lowest $s$-wave state of size
$\sim l_\mathrm{ho}$ could be suppressed, for example, by controlling
the system parameters.

We conclude this paper by discussing a simple analogy between a tower of
our confinement-induced molecules and Kaluza-Klein modes in
extra-dimension models~\cite{Maartens:2010ar}.  Suppose our 3D world has
one compact extra dimension with an extent of $L$.  Then the momentum of
a 4D particle in the extra fourth direction is quantized as
$p_4=2\pi\hbar n/L$.  Because we live in 3D, such a particle can be
viewed as a tower of ``new particles'' with the quantized masses
$m_n=2\pi\hbar n/cL$.  These new particles are called Kaluza-Klein modes
and could be observed by colliding high-energy particles
$>2\pi\hbar c/L$ at the Large Hadron Collider.

Now in our case, for example, in the 2D-3D mixed dimensions, the motion
of a 3D molecule in the ``extra'' $z$ direction is quantized as in
Eq.~(\ref{eq:approx_2D}) because of the confinement potential.  We can
view such quantized energy levels as a tower of new ``Feshbach
molecules'' analogous to the Kaluza-Klein modes.  Unlike in
extra-dimension models, our particle is a composite molecule and we can
control its binding energy to shift the energies of the tower of new
molecules.  When each of them crosses the scattering threshold, the
resonance occurs.  A series of such resonances caused by the
``Kaluza-Klein'' tower of confinement-induced molecules has been
investigated in this paper and observed in the cold-atom
experiment~\cite{Nishida:2008kr,Lamporesi:2010}.

\acknowledgments
The authors thank Yvan Castin, Eugene Demler, David Pekker, Andrey
Turlapov, Fa Wang, Zhenhua Yu, and Martin Zwierlein for valuable
discussions and Jacopo Catani, Giacomo Lamporesi, and Francesco Minardi
for providing the experimental data.  Y.\,N.\ was supported by MIT
Pappalardo Fellowship in Physics and the DOE Office of Nuclear Physics
under Grant No.\ DE-FG02-94ER40818.  S.\,T.\ thanks the Institute for
Advanced Study in Tsinghua University for hospitality when this work was
near completion.

\onecolumngrid\appendix

\section{0D-3D mixed dimensions \label{app:0D-3D}}
Here we present some details of our method to solve integral equations
derived in Sec.~\ref{sec:0D-3D}.  The central issue is the evaluation of
the regular part of the Green's function defined in
Eq.~(\ref{eq:regular_0D}).  When
$E-\frac32\omega\equiv\frac{k^2}{2m_B}\leq0$, it is useful to represent
$\G$ as
\begin{equation}
 \begin{split}
  \G(\r;\r') &= -\int_0^\infty\!d\tau\,e^{\frac{k^2}{2m_B}\tau}
  \left(\frac{m_A\omega\,e^{\omega\tau}}{2\pi\sinh\omega\tau}\right)^{3/2}
  \left(\frac{m_B}{2\pi\tau}\right)^{3/2}\exp\!\left[-\frac{m_A\omega}2
  \frac{(\r+\r')^2\cosh\omega\tau-2\r\cdot\r'}{\sinh\omega\tau}
  -\frac{m_B}{2\tau}(\r-\r')^2\right] \\
  &\quad + \int_0^\infty\!d\tau
  \left(\frac{m_{AB}}{2\pi\tau}\right)^{3/2}\delta(\r-\r').
 \end{split}
\end{equation}
Its partial-wave projection gives
\begin{equation}
 \begin{split}
  \G_\ell(r;r') &\equiv \frac12\int_{-1}^1\!d\cos\theta\,
  \G(\r;\r')\,P_\ell(\cos\theta) \\
  &= -\int_0^\infty\!d\tau\,e^{\frac{k^2}{2m_B}\tau}
  \left(\frac{m_A\omega\,e^{\omega\tau}}{2\pi\sinh\omega\tau}\right)^{3/2}
  \left(\frac{m_B}{2\pi\tau}\right)^{3/2}
  i^{-l}j_l\!\left[i\left(\frac{m_A\omega}{\sinh\omega\tau}
  +\frac{m_B}\tau\right)rr'\right] \\ &\quad \times
  \exp\!\left[-\left(\frac{m_A\omega\cosh\omega\tau}{\sinh\omega\tau}
  +\frac{m_B}\tau\right)\frac{r^2+r'^2}2\right]
  + \int_0^\infty\!d\tau\left(\frac{m_{AB}}{2\pi\tau}\right)^{3/2}
  \frac{\delta(r-r')}{4\pi r^2}.
 \end{split}
\end{equation}

We now evaluate the matrix elements of $\G_\ell$ with respect to the
eigenfunctions of 3D harmonic oscillator with orbital angular momentum
$\ell$:
\begin{equation}
 \phi_n^{(\ell)}(r) \equiv \frac1{l_\mathrm{ho}^{3/2}}
  \sqrt{\frac{n!}{2\pi\left(n+\ell+\frac12\right)!}}\,
  e^{-r^2/(2l_\mathrm{ho}^2)}\left(\frac{r}{l_\mathrm{ho}}\right)^\ell
  L_n^{\left(\ell+\frac12\right)}(r^2/l_\mathrm{ho}^2),
\end{equation}
which form an orthonormal basis:
\begin{equation}
 \int\!d\r\,\phi_n^{(\ell)}(r)\phi_{n'}^{(\ell)}(r)
 = \int_0^\infty\!4\pi r^2dr\,
 \phi_n^{(\ell)}(r)\phi_{n'}^{(\ell)}(r) = \delta_{nn'}.
\end{equation}
A lengthy but straightforward calculation leads to
\begin{equation}
 \begin{split}
  M_{ij}^{(\ell)} &\equiv \frac{2\pi l_\mathrm{ho}}{m_{AB}}
  \int\!d\r d\r'\phi_i^{(\ell)}(r)\G_\ell(r;r')\phi_j^{(\ell)}(r') \\
  &= \frac{-1}{4\sqrt\pi}\sqrt{\frac{m_B}{m_A}}\frac{m_B}{m_{AB}}
  \sqrt{\binom{i+l+\frac12}{i}\binom{j+l+\frac12}{j}}
  \int_0^\infty\!\frac{\omega d\tau}{(\omega\tau)^{3/2}}\,
  e^{\frac{k^2}{2m_B}\tau}
  \left(\frac{e^{\omega\tau}}{\sinh\omega\tau}\right)^{3/2}
  Y^l\left(\frac1{X^2-Y^2}\right)^{l+\frac32} \\
  &\quad \times \left(1-\frac{X}{X^2-Y^2}\right)^{i+j}
  {}_{2\!}F_1\!\left[-i,-j;l+\frac32;
  \left(\frac{Y}{X^2-X-Y^2}\right)^2\right]
  + \delta_{ij}\sqrt{\frac{m_{AB}}{2\pi m_A}}
  \int_0^\infty\!\frac{\omega d\tau}{(\omega\tau)^{3/2}},
 \end{split}
\end{equation}
where ${}_{2\!}F_1$ is the hypergeometric function and
\begin{equation}
 X \equiv \frac12
  \left(\frac1{\tanh\omega\tau}+\frac{m_B}{m_A\omega\tau}+1\right)
  \qquad\text{and}\qquad
  Y \equiv \frac12
  \left(\frac1{\sinh\omega\tau}+\frac{m_B}{m_A\omega\tau}\right).
\end{equation}
Expanding $\chi_\ell$ in terms of $\phi_n^{(\ell)}$, the integral
equations in Sec.~\ref{sec:0D-3D} reduce to linear algebraic equations
that have been solved numerically.

\section{1D-3D mixed dimensions \label{app:1D-3D}}
Here we present some details of our method to solve integral equations
derived in Sec.~\ref{sec:1D-3D}.  The central issue is the evaluation of
the regular part of the Green's function defined in
Eq.~(\ref{eq:regular_1D}).  When $E-\omega\equiv\frac{k^2}{2m_B}\leq0$,
it is useful to represent $\G$ as
\begin{equation}
 \begin{split}
  \G(\bm\rho;\bm\rho') &= -\int_0^\infty\!d\tau\,e^{\frac{k^2}{2m_B}\tau}
  \frac{m_A\omega\,e^{\omega\tau}}{2\pi\sinh\omega\tau}
  \frac{m_B}{2\pi\tau}\sqrt\frac{m_{AB}}{2\pi\tau}
  \exp\!\left[-\frac{m_A\omega}2\frac{(\bm\rho+\bm\rho')^2
  \cosh\omega\tau-2\bm\rho\cdot\bm\rho'}{\sinh\omega\tau}
  -\frac{m_B}{2\tau}(\bm\rho-\bm\rho')^2\right] \\
  &\quad + \int_0^\infty\!d\tau
  \left(\frac{m_{AB}}{2\pi\tau}\right)^{3/2}\delta(\bm\rho-\bm\rho').
 \end{split}
\end{equation}
Its partial-wave projection gives
\begin{equation}
 \begin{split}
  \G_m(\rho;\rho') &\equiv \frac1\pi\int_0^\pi\!d\varphi\,
  \G(\bm\rho;\bm\rho')\cos(m\varphi) \\
  &= -\int_0^\infty\!d\tau\,e^{\frac{k^2}{2m_B}\tau}
  \frac{m_A\omega\,e^{\omega\tau}}{2\pi\sinh\omega\tau}
  \frac{m_B}{2\pi\tau}\sqrt{\frac{m_{AB}}{2\pi\tau}}\,
  I_m\!\left[\left(\frac{m_A\omega}{\sinh\omega\tau}
  +\frac{m_B}\tau\right)\rho\rho'\right] \\ &\quad \times
  \exp\!\left[-\left(\frac{m_A\omega\cosh\omega\tau}{\sinh\omega\tau}
  +\frac{m_B}\tau\right)\frac{\rho^2+\rho'^2}2\right]
  + \int_0^\infty\!d\tau\left(\frac{m_{AB}}{2\pi\tau}\right)^{3/2}
  \frac{\delta(\rho-\rho')}{2\pi\rho}.
 \end{split}
\end{equation}

We now evaluate the matrix elements of $\G_m$ with respect to the
eigenfunctions of 2D harmonic oscillator with magnetic quantum number
$m$:
\begin{equation}
 \phi_n^{(m)}(\rho) \equiv \frac1{l_\mathrm{ho}}
  \sqrt{\frac{n!}{\pi(n+m)!}}\,e^{-\rho^2/(2l_\mathrm{ho}^2)}
  \left(\frac{\rho}{l_\mathrm{ho}}\right)^m
  L_n^{(m)}(\rho^2/l_\mathrm{ho}^2),
\end{equation}
which form an orthonormal basis:
\begin{equation}
 \int\!d\bm\rho\,\phi_n^{(m)}(\rho)\phi_{n'}^{(m)}(\rho)
 = \int_0^\infty\!2\pi \rho\,d\rho\,
 \phi_n^{(m)}(\rho)\phi_{n'}^{(m)}(\rho) = \delta_{nn'}.
\end{equation}
A lengthy but straightforward calculation leads to
\begin{equation}
 \begin{split}
  M_{ij}^{(m)} &\equiv \frac{2\pi l_\mathrm{ho}}{m_{AB}}
  \int\!d\bm\rho d\bm\rho'
  \phi_i^{(m)}(\rho)\G_m(\rho;\rho')\phi_j^{(m)}(\rho') \\
  &= \frac{-1}{2\sqrt{2\pi}}\frac{m_B}{\sqrt{m_Am_{AB}}}
  \sqrt{\binom{i+m}{i}\binom{j+m}{j}}
  \int_0^\infty\!\frac{\omega d\tau}{(\omega\tau)^{3/2}}\,
  e^{\frac{k^2}{2m_B}\tau}\frac{e^{\omega\tau}}{\sinh\omega\tau}
  \,Y^m\left(\frac1{X^2-Y^2}\right)^{m+1} \\
  &\quad \times \left(1-\frac{X}{X^2-Y^2}\right)^{i+j}
  {}_{2\!}F_1\!\left[-i,-j;m+1;
  \left(\frac{Y}{X^2-X-Y^2}\right)^2\right]
  + \delta_{ij}\sqrt{\frac{m_{AB}}{2\pi m_A}}
  \int_0^\infty\!\frac{\omega d\tau}{(\omega\tau)^{3/2}}.
 \end{split}
\end{equation}
Expanding $\chi_m$ in terms of $\phi_n^{(m)}$, the integral equations in
Sec.~\ref{sec:1D-3D} reduce to linear algebraic equations that have been
solved numerically.

\section{2D-3D mixed dimensions \label{app:2D-3D}}
Here we present some details of our method to solve integral equations
derived in Sec.~\ref{sec:2D-3D}.  The central issue is the evaluation of
the regular part of the Green's function defined in
Eq.~(\ref{eq:regular_2D}).  When
$E-\frac12\omega\equiv\frac{k^2}{2m_B}\leq0$, it is useful to represent
$\G$ as
\begin{equation}
 \begin{split}
  \G(z;z') &= -\int_0^\infty\!d\tau\,e^{\frac{k^2}{2m_B}\tau}
  \sqrt\frac{m_A\omega\,e^{\omega\tau}}{2\pi\sinh\omega\tau}
  \sqrt\frac{m_B}{2\pi\tau}\,\frac{m_{AB}}{2\pi\tau}
  \exp\!\left[-\frac{m_A\omega}2\frac{(z+z')^2\cosh\omega\tau-2zz'}{\sinh\omega\tau}
  -\frac{m_B}{2\tau}(z-z')^2\right] \\
  &\quad + \int_0^\infty\!d\tau
  \left(\frac{m_{AB}}{2\pi\tau}\right)^{3/2}\delta(z-z').
 \end{split}
\end{equation}
Its parity projection gives
\begin{equation}
 \begin{split}
  \G_\pm(z;z') &\equiv \frac{\G(z;z')\pm\G(z;-z')}2 \\
  &= -\int_0^\infty\!d\tau\,e^{\frac{k^2}{2m_B}\tau}
  \sqrt{\frac{m_A\omega\,e^{\omega\tau}}{2\pi\sinh\omega\tau}}
  \sqrt{\frac{m_B}{2\pi\tau}}\,\frac{m_{AB}}{2\pi\tau}
  \begin{Bmatrix}
   \cosh \\ \sinh
  \end{Bmatrix}
  \!\left[\left(\frac{m_A\omega}{\sinh\omega\tau}
  +\frac{m_B}\tau\right)zz'\right] \\ &\quad \times
  \exp\!\left[-\left(\frac{m_A\omega\cosh\omega\tau}{\sinh\omega\tau}
  +\frac{m_B}\tau\right)\frac{z^2+z'^2}2\right]
  + \int_0^\infty\!d\tau\left(\frac{m_{AB}}{2\pi\tau}\right)^{3/2}
  \frac{\delta(z-z')\pm\delta(z+z')}2.
 \end{split}
\end{equation}

We now evaluate the matrix elements of $\G_\pm$ with respect to the
eigenfunctions of 1D harmonic oscillator:
\begin{equation}
 \phi_n(z) \equiv \frac1{\sqrt{l_\mathrm{ho}}}
  \frac{\pi^{-1/4}}{\sqrt{2^nn!}}\,e^{-z^2/(2l_\mathrm{ho}^2)}
  H_n(z/l_\mathrm{ho}),
\end{equation}
which form an orthonormal basis:
\begin{equation}
 \int\!dz\,\phi_n^*(z)\phi_{n'}(z) = \delta_{nn'}.
\end{equation}
A lengthy but straightforward calculation leads to
\begin{equation}
 \begin{split}
  M_{ij}^{(+)} &\equiv \frac{2\pi l_\mathrm{ho}}{m_{AB}}
  \int\!dzdz'\phi_i(z)\G_+(z;z')\phi_j(z') \\
  &= \frac1{2\sqrt\pi}\sqrt{\frac{m_B}{m_A}}
  \frac{(-1)^{\frac{i}2+\frac{j}2+1}}
  {\left(\frac{i}2\right)!\left(\frac{j}2\right)!}
  \sqrt\frac{i!\,j!}{2^{i+j}}
  \int_0^\infty\!\frac{\omega d\tau}{(\omega\tau)^{3/2}}\,
  e^{\frac{k^2}{2m_B}\tau}\sqrt{\frac{e^{\omega\tau}}{\sinh\omega\tau}}
  \,\left(\frac1{X^2-Y^2}\right)^{1/2} \\
  &\quad \times \left(1-\frac{X}{X^2-Y^2}\right)^{\frac{i}2+\frac{j}2}
  {}_{2\!}F_1\!\left[-\frac{i}2,-\frac{j}2;\frac12;
  \left(\frac{Y}{X^2-X-Y^2}\right)^2\right]
  + \delta_{ij}\sqrt{\frac{m_{AB}}{2\pi m_A}}
  \int_0^\infty\!\frac{\omega d\tau}{(\omega\tau)^{3/2}}
 \end{split}
\end{equation}
for $i,j=\text{even integers}$ and
\begin{equation}
 \begin{split}
  M_{ij}^{(-)} &\equiv \frac{2\pi l_\mathrm{ho}}{m_{AB}}
  \int\!dzdz'\phi_i(z)\G_-(z;z')\phi_j(z') \\
  &= \frac1{\sqrt\pi}\sqrt{\frac{m_B}{m_A}}
  \frac{(-1)^{\frac{i}2+\frac{j}2}}
  {\left(\frac{i-1}2\right)!\left(\frac{j-1}2\right)!}
  \sqrt\frac{i!\,j!}{2^{i+j}}
  \int_0^\infty\!\frac{\omega d\tau}{(\omega\tau)^{3/2}}\,
  e^{\frac{k^2}{2m_B}\tau}\sqrt{\frac{e^{\omega\tau}}{\sinh\omega\tau}}
  \,Y\left(\frac1{X^2-Y^2}\right)^{3/2} \\
  &\quad \times \left(1-\frac{X}{X^2-Y^2}\right)^{\frac{i}2+\frac{j}2-1}
  {}_{2\!}F_1\!\left[\frac{1-i}2,\frac{1-j}2;\frac32;
  \left(\frac{Y}{X^2-X-Y^2}\right)^2\right]
  + \delta_{ij}\sqrt{\frac{m_{AB}}{2\pi m_A}}
  \int_0^\infty\!\frac{\omega d\tau}{(\omega\tau)^{3/2}}
 \end{split}
\end{equation}
for $i,j=\text{odd integers}$.  Expanding $\chi_\pm$ in terms of
$\phi_n$, the integral equations in Sec.~\ref{sec:2D-3D} reduce to
linear algebraic equations that have been solved numerically.

\twocolumngrid

\end{document}